 \documentclass[preprint,nofootinbib,aps]{revtex4}
 \usepackage{graphicx}
 \usepackage{color}
 \usepackage{amssymb}
 \usepackage[colorlinks]{hyperref}

\begin{document}
 \title{Study of the $B_{s}$ ${\to}$ ${\phi}f_{0}(980)$ ${\to}$
   ${\phi}\,{\pi}^{+}{\pi}^{-}$ decay \\ with perturbative QCD approach}
 \author{Na Wang}
 \email[]{wangna05001@126.com}
 \affiliation{Institute of Particle and Nuclear physics, \\
              Henan Normal University, Xinxiang 453007, China}
 \author{Qin Chang}
 \affiliation{Institute of Particle and Nuclear physics, \\
              Henan Normal University, Xinxiang 453007, China}
 \author{Yueling Yang}
 \affiliation{Institute of Particle and Nuclear physics, \\
              Henan Normal University, Xinxiang 453007, China}
 \author{Junfeng Sun}
 \affiliation{Institute of Particle and Nuclear physics, \\
              Henan Normal University, Xinxiang 453007, China}
 \begin{abstract}
 The rare cascade $B_{s}$ ${\to}$ ${\phi}f_{0}(980)$ ${\to}$
 ${\phi}\,{\pi}^{+}{\pi}^{-}$ decay is studied with the
 perturbative QCD approach based on the formula for the
 quasi two-body decay, where the two-pion pair originates
 from the $S$-wave resonant $f_{0}(980)$ state.
 It is found that with the introduction of the nonperturbative
 two-pion distribution amplitudes and the Flatt\'{e}
 parameterization of the scalar form factor for the $f_{0}(980)$
 resonance, the branching ratio in the mass range
 400 MeV $<$ $m({\pi}^{+}{\pi}^{-})$ $<$ 1600 MeV
 is ${\cal B}_{\rm theo}(B_{s}^{0}\,
 {\to}\,{\phi}\,f_{0}(980)\,{\to}\,{\phi}\,{\pi}^{+}\,{\pi}^{-})$
 $=$ $[1.31_{-0.31}^{+0.40}(a_{{\pi}{\pi}}){}_{-0.16}^{+0.19}(m_{b})
 {}_{-0.09}^{+0.10}(\text{CKM})]{\times}10^{-6}$, where the uncertainties
 come from the parameter $a_{{\pi}{\pi}}$ of the two-pion distribution
 amplitudes, the $b$ quark mass $m_{b}$, the Cabibbo-Kobayashi-Maskawa (CKM)
 factors, respectively. This result agrees with the recent LHCb measurement
 within uncertainties, ${\cal B}_{\rm exp}(B_{s}^{0}\,{\to}\,{\phi}\,
 f_{0}(980)\,{\to}\,{\phi}\,{\pi}^{+}\,{\pi}^{-})$ $=$
 $(1.12{\pm}0.16^{+0.09}_{-0.08}{\pm}0.11){\times}10^{-6}$,
 where the errors are statistical, systematic and from the
 normalization, respectively.
 \end{abstract}
 \maketitle

 \section{introduction}
 \label{sec01}
  With the great progress and good performance of the Belle, BaBar and LHCb
  experiments, many three-body nonleptonic $B$ meson weak decay channels are
  accessible and have been measured \cite{pdg}. Recently, based on the $3\,fb^{-1}$
  $pp$ collision data recorded by the LHCb detector, the three-body nonleptonic
  decay $B_{s}^{0}$ ${\to}$ ${\phi}\,{\pi}^{+}{\pi}^{-}$
  was investigated with the requirements on the ${\pi}^{+}{\pi}^{-}$ invariant mass
  in the range 400 MeV $<$ $m({\pi}{\pi})$ $<$ 1600 MeV, then an analysis of the
  $m({\pi}{\pi})$ spectrum including the $S$-, $P$-, and $D$-wave amplitudes was
  further performed to study the possible resonant contributions \cite{prd95.012006}.
  Some prominent maxima in the $m({\pi}{\pi})$ spectrum are observed around the
  ${\rho}(770)$, $f_{0}(980)$, $f_{2}(1270)$ and $f_{0}(1500)$ resonant regions.
  One of the formal public announcement of the LHCb Collaboration is that \cite{prd95.012006}
  the three-body sequential rare decay $B_{s}^{0}$ ${\to}$ ${\phi}f_{0}(980)$
  ${\to}$ ${\phi}{\pi}^{+}{\pi}^{-}$ was first observed with a statistical
  significance of $8\,{\sigma}$ and the branching fraction of
  \begin{equation}
  {\cal B}(B_{s}^{0}\,{\to}\,{\phi}\,f_{0}(980)\,{\to}\,{\phi}\,{\pi}^{+}\,{\pi}^{-})
   \, =\, (1.12{\pm}0.16^{+0.09}_{-0.08}{\pm}0.11){\times}10^{-6}
  \label{eq:br-exp-01},
  \end{equation}
  where the errors are statistical, systematic and from the normalization,
  respectively.

  Although it is still a controversial issue whether the isospin-singlet
  particle $f_{0}(980)$ should be regarded as the conventional $q\bar{q}$
  meson, or the exotic tetraquark $q\bar{q}q\bar{q}$ state, or the
  meson-meson $K\bar{K}$ molecule, it is usually suggested that the
  unflavored scalar $f_{0}(980)$ meson has a substantial $s\bar{s}$ component,
  and decays dominantly into the ${\pi}{\pi}$ final states \cite{pdg}.
  Therefore, on the one hand, the $B_{s}^{0}$ ${\to}$ ${\phi}f_{0}(980)$
  ${\to}$ ${\phi}\,{\pi}^{+}{\pi}^{-}$ decay is interesting and helpful
  to explore the compositive structure of the $f_{0}(980)$;
  on the other hand, the importance of the $B_{s}^{0}$ ${\to}$
  ${\phi}f_{0}(980)$ ${\to}$ ${\phi}\,{\pi}^{+}{\pi}^{-}$ decay is obvious,
  {\em i.e.}, this decay is induced by the flavor-changing-neutral-current
  (FCNC) $\bar{b}$ ${\to}$ $\bar{s}s\bar{s}$ process at the elementary
  particle level within the Standard model (SM), which is absolutely
  forbidden at the tree level by the Cabibbo-Kobayashi-Maskawa (CKM)
  quark-mixing mechanism within SM but sensitive to the new physics
  effects beyond SM.

  Along with the experimental advances, the theoretical research on the
  three-body nonleptonic $B$ weak decay is really necessary.
  Although there exist some attractive QCD-inspired phenomenological
  methods to deal with the two-body nonleptonic $B$ decays, such as
  the perturbative QCD (PQCD) approach \cite{plb348.597,prd52.3958,prd55.5577,
  prd56.1615,plb504.6,prd63.054008,prd63.074006,prd63.074009,epjc23.275},
  the QCD factorization (QCDF) approach \cite{prl83.1914,npb591.313,
  npb606.245,plb488.46,plb509.263,prd64.014036,npb774.64,prd77.074013},
  and so on, the theoretical description of the three-body nonleptonic
  $B$ decays is still in the early stage of modeling.
  This is not surprising because that the more hadrons participated,
  the more intricate the interferences among different contributions
  (such as the possible resonances and final state interactions) will
  certainly become. Moreover, for the three-body hadronic decays,
  the kinematical configurations will vary from region to region in
  the Dalitz plot, and in principle correspond to different dynamical
  components and theoretical treatments with special scales.
  The resonant contributions are entirely engulfed by the blurry
  background clouds, so any phenomenological parametrization and
  interpretations of the resonant structures are process- and
  model-dependent.
  The effective separations between the perturbative and nonperturbative
  contributions to the three-body nonleptonic $B$ decays will be much
  more complicated, which is by no means trivial.
  However at the same time, the phase space distributions make the
  theoretical calculation of three-body nonleptonic $B$ meson decays
  to be very meaningful for exploring some fresh and potentially important
  information, such as the natures and effects of possible resonances,
  the energy dependence of observables, the local $CP$ asymmetry
  distributions in the Dalitz plot, and so on.
  In the past years, there were plenty of theoretical studies of the
  three-body nonleptonic $B$ meson decays, such as
  Refs.\cite{prd39.3346,prd44.1454,prd52.6356,prd68.015004,plb564.90,
  prd72.094031,prd75.014002,plb726.337,plb727.136,plb728.206,prd89.074043,
  prd84.034040,prd84.034041,prd84.056002,prd85.016010,prl90.061802,
  prd74.051301,ijmpa29.1450011,plb728.579,prd91.014029}
  based on SU(3) relations,
  Refs.\cite{prd52.5354,prd60.054029,plb447.313,plb539.67,prd69.114020,
  prd70.034033,prd65.094004,prd66.054015,ijmpa23.3229,prd72.094003,prd76.094006,
  prd88.114014,prd89.074025,prd89.094007,prd94.094015,plb665.30}
  based on both heavy quark effective theory and chiral perturbation
  theory,
  Refs.\cite{epjc31.215,jhep1602.009,prd65.034003,prd66.054004,prd67.034012,
  prd70.034032,epjc33.s253,epja50.122,cjp93.339,prd89.095026,prd95.036013,
  prd96.113003,prd62.036001,prd62.114011,prl86.216,prd87.076007}
  with factorization approach,
  Refs.\cite{jpg31.199,prd83.014002,prd90.034014,ahep2014.785648,jhep1710.117,
  npb899.247,prd93.116008,aplb42.2013,plb622.207,prd74.114009,prd79.094005,
  prd81.094033,plb699.102,plb737.201,ps89.095301,ahep2014.451613}
  with the QCDF approach,
  Refs.\cite{plb561.258,prd70.054006,npa930.117,prd89.074031,prd91.094024,
  epjc76.675,plb763.29,cpc41.083105,npb923.54,npb924.745,prd95.056008,
  prd96.036014,prd96.093011,epjc77.199,prd94.034040,epjc77.518}
  with the PQCD approach.

  Both the PQCD and QCDF approaches have been widely employed in the two-body
  nonleptonic $B$ meson decays in recent years.
  In Ref.\cite{plb561.258}, Chen and Li attempted to generalize the PQCD
  approach to the three-body nonleptonic $B^{+}$ ${\to}$ $K^{+}{\pi}^{+}{\pi}^{-}$
  decay for the particular configuration topologies where the kinematics is very
  similar to a two-body decay.
  In this paper, we shall follow the method of Ref.\cite{plb561.258} to investigate
  the $B_{s}^{0}$ ${\to}$ ${\phi}f_{0}(980)$ ${\to}$ ${\phi}\,{\pi}^{+}{\pi}^{-}$
  decay with the PQCD approach.
  The overall layout of this paper is as follows.
  The theoretical framework and the amplitudes for the $B_{s}^{0}$ ${\to}$
  ${\phi}f_{0}(980)$ ${\to}$ ${\phi}\,{\pi}^{+}{\pi}^{-}$ decay are
  elaborated in section \ref{sec02}.
  The numerical results and discussion are presented in Section \ref{sec03}.
  The last section is a short summary.

  \section{theoretical framework}
  \label{sec02}
  \subsection{The effective Hamiltonian}
  \label{sec0201}
  The nonleptonic weak decays of the $B$ mesons involve three fundamental scales,
  including the weak interaction scale $M_{W}$, the $b$ quark mass scale $m_{b}$,
  and the QCD characteristic scale ${\Lambda}_{\rm QCD}$, which are strongly
  ordered: $M_{W}$ ${\gg}$ $m_{b}$ ${\gg}$ ${\Lambda}_{\rm QCD}$.
  To deal with the multi-scale problems, one usually has to resort to the
  effective theory approximation.
  Using the operator product expansion and the renormalization group (RG) equation,
  the low energy effective Hamiltonian for the FCNC $B_{s}^{0}$ ${\to}$
  ${\phi}f_{0}(980)$ ${\to}$ ${\phi}\,{\pi}^{+}{\pi}^{-}$ decay within
  the SM can be written as \cite{rmp68.1125}:
  \begin{equation}
 {\cal H}_{\rm eff}\, =\,
  \frac{G_{F}}{\sqrt{2}}\, \Big(V_{ub}^{\ast}\,V_{us}+V_{cb}^{\ast}\,V_{cs}\Big)\,
  \sum_{i=3}^{10}\,C_{i}({\mu})\,Q_{i}({\mu})\, +\, {\rm h.c.}
  \label{eq:Heff},
  \end{equation}
  where the Fermi coupling constant $G_{F}$ ${\simeq}$ $1.166{\times}10^{-5}$
  ${\rm GeV}^{-2}$ \cite{pdg}. $V_{ub}^{\ast}\,V_{us}$ and $V_{cb}^{\ast}\,V_{cs}$
  are the CKM factors. The scale ${\mu}$ separates the effective Hamiltonian into
  two distinct parts: the Wilson coefficients $C_{i}$ and the local four-quark
  operators $Q_{i}$.

  The expressions of the operators $Q_{i}$ are written as:
  \begin{eqnarray}
  Q_{3} &=& \sum_{q}\,
        (\bar b_{\alpha}\,s_{\alpha})_{V-A}\,
        (\bar q_{\beta}\,q_{\beta})_{V-A}, \,
  \qquad \quad \
  Q_{4}\, =\, \sum_{q}\,
        (\bar b_{\alpha}\,s_{\beta})_{V-A}\,
        (\bar q_{\beta}\,q_{\alpha})_{V-A},
  \label{q3-q4} \\
  Q_{5} &=& \sum_{q}\,
        (\bar b_{\alpha}\,s_{\alpha})_{V-A}\,
        (\bar q_{\beta}\,q_{\beta})_{V+A}, \,
  \qquad \quad \
  Q_{6}\, =\, \sum_{q}\,
        (\bar b_{\alpha}\,s_{\beta})_{V-A}\,
        (\bar q_{\beta}\,q_{\alpha})_{V+A},
  \label{q5-q6}  \\
  Q_{7} &=& \frac{3}{2}\,\sum_{q}\,Q_{q}\,
       (\bar b_{\alpha}\,s_{\alpha})_{V-A}\,
       (\bar q_{\beta}\,q_{\beta})_{V+A},
  \quad
  Q_{8}\, =\, \frac{3}{2}\,\sum_{q}\,Q_{q}\,
       (\bar b_{\alpha}\,s_{\beta})_{V-A}\,
       (\bar q_{\beta}\,q_{\alpha})_{V+A},
  \label{q7-q8}  \\
  Q_{9} &=& \frac{3}{2}\,\sum_{q}\,Q_{q}\,
       (\bar b_{\alpha}\,s_{\alpha})_{V-A}\,
       (\bar q_{\beta}\,q_{\beta})_{V-A},
  \quad
  Q_{10}\, =\, \frac{3}{2}\,\sum_{q}\,Q_{q}\,
       (\bar b_{\alpha}\,s_{\beta})_{V-A}\,
       (\bar q_{\beta}\,q_{\alpha})_{V-A},
  \label{q9-q10}
  \end{eqnarray}
  where $Q_{3,...,6}$ and $Q_{7,...,10}$ are called as the QCD and
  electroweak penguin operators, respectively.
  ${\alpha}$ and ${\beta}$ are color indices.
  $q$ denotes all the active quark at the scale of ${\cal{O}}(m_{b})$,
  {\em i.e.}, $q$ $=$ $u$, $d$, $s$, $c$, $b$.
  $Q_{q}$ is the electric charge of quark $q$ in the unit of
  ${\vert}e{\vert}$.

  The operators $Q_{i}$ govern the dynamics of the $B$ meson weak decay.
  The coupling strength of the effective interactions among four quarks
  of the operators $Q_{i}$ is proportionate to the Wilson coefficients
  $C_{i}$. The physical contributions from the scale higher than ${\mu}$
  are summarized in the Wilson coefficients $C_{i}$, while
  the physical contributions from the scale lower than ${\mu}$ are
  incorporated into the hadronic matrix elements (HMEs) where the operators
  $Q_{i}$ are sandwiched between the initial and final hadron states.
  The Wilson coefficients $C_{i}$ are process independent and computable
  order by order with the RG improved pertrubative theory as long as the
  scale ${\mu}$ is not too small. The expressions of the Wilson coefficients
  $C_{i}$ including the next-to-leading order corrections can be found in
  Ref.\cite{rmp68.1125}.
  The HMEs describe the transition from the quarks of the operators $Q_{i}$
  to the participating hadrons.
  The operators $Q_{i}$ comprise of four quarks at the local interaction
  point, the initial and final states are hadronic states. The transition
  between the quarks and hadrons necessarily involves the hadronization
  and other rescattering effects.
  Due to the low-energy long-distance QCD effects and the entanglement
  of nonperturbative and perturbative contributions, the main obstacles
  of the calculation of the nonleptonic $B$ decays is how to properly
  evaluate the HMEs of the local four-quark operators.

  \subsection{Hadronic matrix element}
  \label{sec0202}
  As for the two-body nonleptonic $B$ decays with both the PQCD and QCDF approaches,
  the HMEs are usually written as the convolution of the universal wave functions
  (WFs) or distribution amplitudes (DAs) reflecting the nonperturbative contributions
  with the scattering amplitudes containing perturbative contributions, based on
  the factorization theorem for exclusive processes \cite{plb87.359,prl43.545,
  prd22.2157,plb91.239,plb90.159}. Similarly, the HMEs for the three-body nonleptonic
  $B$ decays could generally be written as:
  \begin{eqnarray}
 {\langle}h_{1}h_{2}h_{3}{\vert}Q_{i}{\vert}B{\rangle} &{\sim}&
 {\int}dk_{1}\,dk_{2}\,dk_{3}\,dk\,
 {\Phi}_{h_{1}}(k_{1})\,{\Phi}_{h_{2}}(k_{2})\,{\Phi}_{h_{3}}(k_{3})\,
 {\Phi}_{B}(k)\,{\cal T}(k_{1},k_{2},k_{3},k)
  \label{hme-01}, \\ \text{or} &{\sim}&
 {\int}dx_{1}\,dx_{2}\,dx_{3}\,dx\,
 {\phi}_{h_{1}}(x_{1})\,{\phi}_{h_{2}}(x_{2})\,{\phi}_{h_{3}}(x_{3})\,
 {\phi}_{B}(x)\,\widetilde{\cal T}(x_{1},x_{2},x_{3},x)
  \label{hme-02},
  \end{eqnarray}
  where ${\Phi}_{h_{i}}(k_{i})$ and ${\phi}_{h_{i}}(x_{i})$ are the
  WFs and DAs for the $h_{i}$ hadron, respectively; $k_{i}$ ($x_{i}$)
  is the momentum (the longitudinal momentum fraction) of the valence
  quark; ${\cal T}$ and $\widetilde{\cal T}$ are the scattering kernels.
  It is assumed that the nonperturbative contributions are contained
  within the WFs and DAs. The DAs are universal.
  The DAs either extracted from experimental data or obtained from
  nonperturbative means could be employed for other processes
  involving the same hadron.
  The scattering kernels, ${\cal T}$ and $\widetilde{\cal T}$, could be
  computed systematically in an expansion in the strong coupling
  ${\alpha}_{s}$ and the power $1/m_{b}$ with the perturbation theory.

  As analyzed in Refs.\cite{npb899.247}, the Dalitz plot for the three-body
  nonleptonice $B$ decays could be divided into different regions with
  distinct kinematic and dynamic properties.
  In the center region of the Dalitz plot, all three final hadrons have a
  large energy and none of them moves collinearly to the others.
  This kinematical configurations have two hard gluons and the perturbative
  calculation of the scattering kernels seems to be applicable.
  Unfortunately, as analyzed in Ref.\cite{plb561.258}, the scattering kernels
  of this region contain two virtual gluons at the lowest order with the PQCD
  approach, which is not practical due to a huge number of Feynman diagrams.
  In addition, the amplitudes for this region are power suppressed with
  respect to the amplitude at the edges. At the edges of the Dalitz plot,
  two hadrons move collinearly or back-to-back, so the three-body decay
  could be approximately regarded as the quasi-two-body decay \cite{npb899.247}.
  The $B_{s}$ ${\to}$ ${\phi}\,{\pi}^{+}{\pi}^{-}$ decay observed by the LHCb
  Collaboration \cite{prd95.012006} with the ${\pi}^{+}{\pi}^{-}$ invariant mass
  less than 1.6 GeV is the case, where the possible ${\pi}^{+}{\pi}^{-}$ resonant
  states show up.
  The three-body $B_{s}$ ${\to}$ ${\phi}\,{\pi}^{+}{\pi}^{-}$ decay could be
  approximated as the quasi two-body $B_{s}$ ${\to}$ ${\phi}\,({\pi}^{+}{\pi}^{-})$
  decay. It seems reasonable to assume that the two-pion pair originates
  from a quark-antiquark state and postulate the validity of factorization
  for this quasi two-body $B_{s}$ decay.
  In this paper, we will follow Ref.\cite{plb561.258} as a hypothesis,
  and write the HMEs for the three-body nonleptonic $B_{s}$ ${\to}$
  ${\phi}\,{\pi}^{+}{\pi}^{-}$ decay as follow.
  \begin{equation}
 {\langle}{\phi}\,{\pi}^{+}{\pi}^{-}{\vert}Q_{i}{\vert}B_{s}{\rangle}\, {\sim}\,
 {\int}dx\,dy\,dz\,
 {\phi}_{B}(x)\,{\phi}_{\phi}(y)\,{\phi}_{{\pi}{\pi}}(z)\,
  \widetilde{\cal T}(x,y,z)
  \label{hme-quasi},
  \end{equation}
  where one new input, the ${\pi}^{+}{\pi}^{-}$ pair DA ${\phi}_{{\pi}{\pi}}$
  parameterizing both the resonant and nonresonant contributions, is introduced
  in order to factorize the HMEs for the three-body decay.
  It is possible to combine the ${\pi}^{+}{\pi}^{-}$ pair with the $S$, $P$, $D$
  waves. The $S$-, $P$- and $D$-wave transition matrix elements between the two-pion
  pair and the vacuum are proportional to the time-like scale, vector, and tensor
  form factors, respectively.
  It is clear that for the sequential $B_{s}^{0}$ ${\to}$ ${\phi}f_{0}(980)$
  ${\to}$ ${\phi}\,{\pi}^{+}{\pi}^{-}$ decay in question, only the $S$ wave
  contribution from the scalar $f_{0}(980)$ meson needs to be considered.

  \subsection{Kinematic variable}
  \label{sec0203}
\begin{figure}[h]
  \includegraphics[width=0.95\textwidth,bb=90 340 520 760]{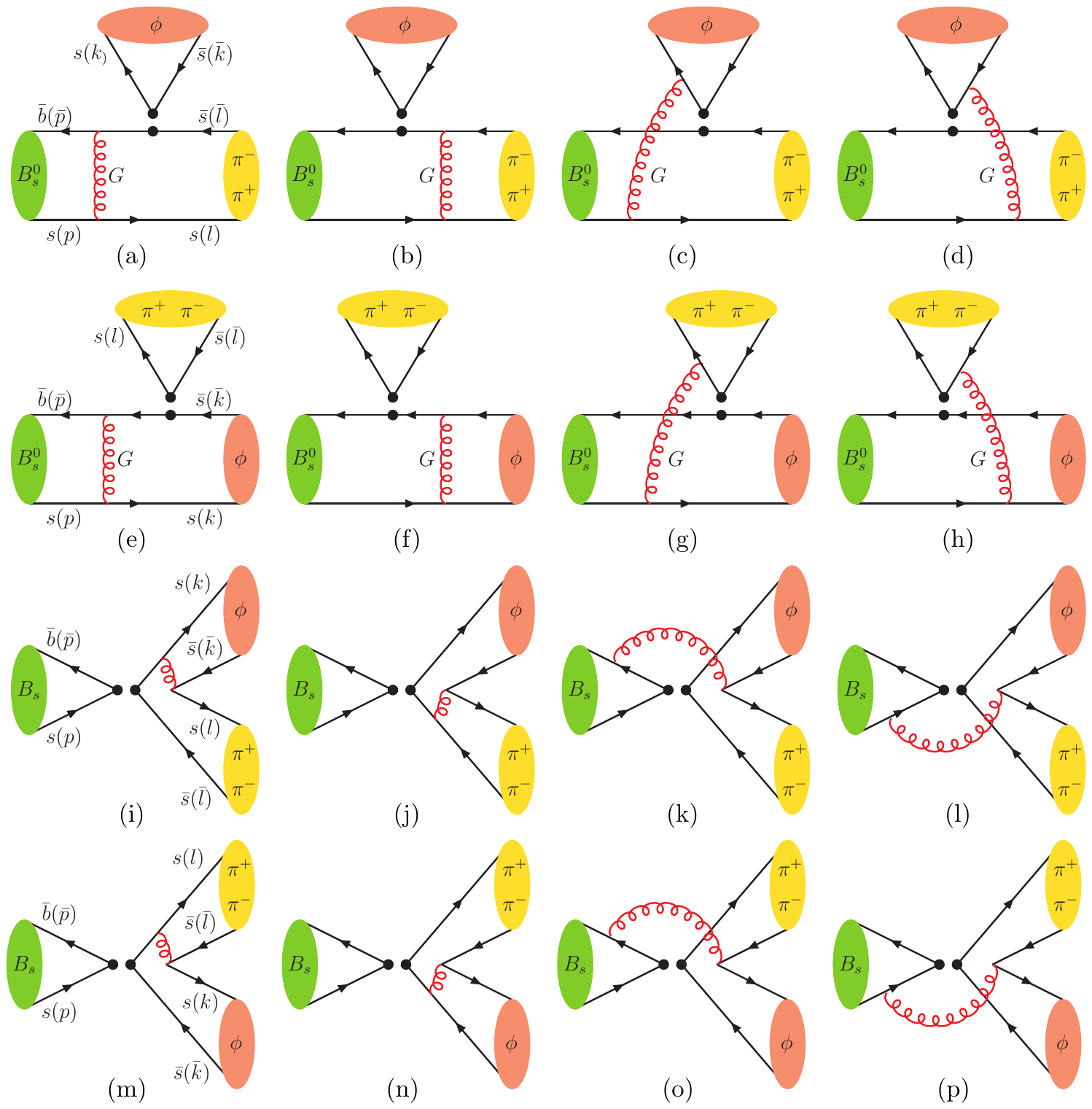}
  \caption{Feynman diagrams for the $B_{s}$ ${\to}$ ${\phi}\,f_{0}(980)$
  ${\to}$ ${\phi}\,\pi^{+}\pi^{-}$ decay with the PQCD approach.}
  \label{fey}
  \end{figure}

  It is convenient to describe the kinematical variables in terms of the light
  cone coordinates. The relations between the four-dimensional space-time
  coordinates ($x^{0}$, $x^{1}$, $x^{2}$, $x^{3}$) $=$ ($t$, $x$, $y$, $z$)
  and the light-cone coordinates ($x^{+}$, $x^{-}$, $x_{\perp}$)
  are defined as $x^{\pm}$ $=$ $(x^{0}{\pm}x^{3})/\sqrt{2}$ and
  $x_{\perp}$ $=$ ($x^{1}$, $x^{2}$).
  The scalar product of two vectors is given by $a{\cdot}b$ $=$
  $a_{\mu}b^{\mu}$ $=$ $a^{+}b^{-}$ $+$ $a^{-}b^{+}$
  $-$ $a_{\perp}{\cdot}b_{\perp}$.
  $n_{+}^{\mu}$ $=$ $(1,0,0)$ and $n_{-}^{\mu}$ $=$ $(0,1,0)$
  are the plus and minus null vectors, respectively \cite{prd96.036010}.
  The momenta of the participating mesons in the rest frame of the $B_{s}$
  meson are defined as follow.
  \begin{equation}
  p_{B_{s}}\, =\, \frac{m_{B_{s}}}{\sqrt{2}}(1,1,0)
  \label{kine-bs},
  \end{equation}
  \begin{equation}
  p_{\phi}\, =\, (p_{\phi}^{+},p_{\phi}^{-},0)
  \label{kine-phi},
  \end{equation}
  \begin{equation}
 {\epsilon}_{\phi}^{\parallel}\, =\, \frac{1}{m_{\phi}}(p_{\phi}^{+},-p_{\phi}^{-},0)
  \label{kine-epsilon},
  \end{equation}
  \begin{equation}
  p_{2{\pi}}\, =\, q\, =\, (q^{-},q^{+},0)
  \label{kine-2pi},
  \end{equation}
  \begin{equation}
  p_{{\pi}^{+}}\, =\, (\bar{\zeta}q^{-},{\zeta}q^{+},+\sqrt{{\zeta}\bar{\zeta}}w)
  \label{kine-pip},
  \end{equation}
  \begin{equation}
  p_{{\pi}^{-}}\, =\, ({\zeta}q^{-},\bar{\zeta}q^{+},-\sqrt{{\zeta}\bar{\zeta}}w)
  \label{kine-pim},
  \end{equation}
  \begin{equation}
  p_{\phi}^{\pm}\, =\, (E_{\phi}\,{\pm}\,p_{\rm cm})/{\sqrt{2}} 
  \label{kine-p1-pm},
  \end{equation}
  \begin{equation}
  q^{\pm}\, =\, (E_{w}\,{\pm}\,p_{\rm cm})/{\sqrt{2}} 
  \label{kine-q-pm},
  \end{equation}
  \begin{equation}
  E_{\phi}\, =\, (m_{B_{s}}^2+m_{\phi}^2-w^2)/(2\,m_{B_{s}}) 
  \label{kine-E1},
  \end{equation}
  \begin{equation}
  E_{w}\, =\, (m_{B_{s}}^2-m_{\phi}^2+w^2)/(2\,m_{B_{s}}) 
  \label{kine-Ew},
  \end{equation}
  \begin{equation}
  p_{\rm cm}\, =\, \frac{\sqrt{[m_{B_{s}}^2-(m_{\phi}+w)^2][m_{B_{s}}^2-(m_{\phi}-w)^2]}}{2\,m_{B_{s}}}
  \label{kine-pcm},
  \end{equation}
  \begin{equation}
  q^2\, =\, (p_{B_{s}}-p_{\phi})^2\, =\, (p_{{\pi}^{+}}+p_{{\pi}^{-}})^2\, =\,w^2
  \label{kine-q2},
  \end{equation}
  where ${\epsilon}_{\phi}^{\parallel}$ is the longitudinal polarization
  vector of the ${\phi}$ meson.  $\bar{\zeta}$ $=$ $1$ $-$ ${\zeta}$.
  The variable ${\zeta}$ ($\bar{\zeta}$) is the ${\pi}^{+}$ (${\pi}^{-}$)
  meson momentum fraction of the ${\pi}^{+}{\pi}^{-}$ meson pair with
  the invariant mass $w$ $=$ $m({\pi}{\pi})$.
  The momenta of the spectator quark of the $B_{s}$ meson and the valence quarks
  of the final states are defined as $p$, $k$ and $l$ (see Fig.\ref{fey} for detail)
  with the longitudinal momentum fraction of $x$, $y$, $z$ and the transverse momentum
  of $p_{T}$, $k_{T}$, $l_{T}$, respectively,
  \begin{equation}
  p\, =\, (xp_{B_{s}}^{+},xp_{B_{s}}^{-},p_{T})
  \label{kine-k},
  \end{equation}
  \begin{equation}
  k\, =\, (yp_{\phi}^{+},yp_{\phi}^{-},k_{T})
  \label{kine-k1},
  \end{equation}
  \begin{equation}
  l\, =\, (zq^{-},zq^{+},l_{T})
  \label{kine-l}.
  \end{equation}

  \subsection{The distribution amplitudes}
  \label{sec0204}
  Within the pQCD framework, the WFs and/or DAs are the essential input parameters.
  Following the notations in Refs.\cite{jhep0703.069,prd65.014007,prd92.074028,
  plb751.171,plb752.322}, the WFs of the $B_{s}$ meson and the longitudinally
  polarized ${\phi}$ meson are defined as:
  \begin{equation}
 {\langle}0{\vert}\bar{b}_{i}(0)s_{j}(z){\vert}B_{s}(p){\rangle}\, =\,
 -\frac{i\,f_{B_{s}}}{4}{\int}d^{4}k\,e^{-ik{\cdot}z}\, \Big\{
  \Big[ \!\!\not{p}\,{\Phi}_{B}^{a}(k) + m_{B_{s}} {\Phi}_{B}^{p}(k) \Big]
 {\gamma}_{5} \Big\}_{ji}
  \label{wf-bs},
  \end{equation}
  \begin{equation}
 {\langle}{\phi}(p,{\epsilon}^{\parallel}){\vert}s_{i}(z)\bar{s}_{j}(0){\vert}0{\rangle}\,=\,
  \frac{1}{4} {\int}_{0}^{1}dk\,e^{ik{\cdot}z}\,\Big\{ \!\!\not{\epsilon}^{\parallel}
  m_{\phi}{\Phi}_{\phi}^{v}(k)+ \!\!\not{\epsilon}^{\parallel}\!\!\!\not{p}\,
  {\Phi}_{\phi}^{t}(k)-m_{\phi}{\Phi}_{\phi}^{s}(k) \Big\}_{ji}
  \label{wf-phi},
  \end{equation}
 where $f_{B_{s}}$ $=$ $227.2{\pm}3.4$ MeV \cite{pdg} is the decay
 constant of the $B_{s}$ meson.
 The WFs of ${\Phi}_{B}^{a}$ and ${\Phi}_{\phi}^{v}$ are twist-2,
 while the WFs of ${\Phi}_{B}^{P}$ and ${\Phi}_{\phi}^{t,s}$ are twist-3.
 By integrating out the transverse momentum from the wave functions,
 one can obtain the corresponding DAs.

 In our calculation, the expressions of the $B_{s}$ DAs are
 \cite{prd92.074028,plb751.171,plb752.322}:
  \begin{equation}
 {\phi}_{B}^{a}(x) \,=\, {\cal N}_{a}\,x\,\bar{x}\,
 {\exp}\Big\{-\frac{1}{8\,{\omega}_{B}} \Big(
  \frac{m_{s}^{2}}{x}+\frac{m_{b}^{2}}{\bar{x}} \Big) \Big\}
  \label{DA-B-a},
  \end{equation}
  \begin{equation}
 {\phi}_{B}^{p}(x) \,=\, {\cal N}_{p}\,
 {\exp}\Big\{-\frac{1}{8\,{\omega}_{B}} \Big(
  \frac{m_{s}^{2}}{x}+\frac{m_{b}^{2}}{\bar{x}} \Big) \Big\}
  \label{DA-B-p},
  \end{equation}
 where $x$ and $\bar{x}$ $=$ $1$ $-$ $x$ are the longitudinal momentum
 fractions of light and heavy quarks, respectively; $m_{b}$ $=$
 $4.78{\pm}0.06$ GeV \cite{pdg} and $m_{s}$ $=$ ${\simeq}$ 0.51 GeV \cite{book.kamal}
 are the mass of the $b$ and $s$ quarks. The parameter ${\omega}_{B}$
 determines the average transverse momentum of partons, and ${\omega}_{B}$
 ${\simeq}$ $m_{i}{\alpha}_{s}$. The parameters ${\cal N}_{a}$ and
 ${\cal N}_{p}$ are the normalization coefficients,
  \begin{equation}
 {\int}_{0}^{1}dx\,{\phi}_{B}^{a, p}(x)\, =\, 1
  \label{DA-B-coe}.
  \end{equation}
 One distinguish feature of the above DAs is the exponential functions,
 which strongly suppress the contribution from the end point of $x$,
 $\bar{x}$ ${\to}$ $0$ and naturally provide the effective truncation
 for the end point and soft contributions. In addition, the exponential
 factors are proportional to the ratio of the parton mass squared $m^{2}_{i}$
 to the momentum fraction $x_{i}$. Hence, the above DAs are generally
 consistent with the ansatz that the momentum fractions are shared among
 the valence quarks according to the quark mass, {\em i.e.}, the light $s$ quark
 carries relatively less momentum fraction in the heavy-light $B_{s}$ meson.

 The expressions of the two-particle DAs of the ${\phi}$ meson are
 \cite{jhep0703.069,prd65.014007}:
  \begin{equation}
 {\phi}_{\phi}^{v}(x) \, =\, 6\,f_{\phi}\,x\,\bar{x}\,
  \Big\{1+a_{2}^{\phi}C_{2}^{3/2}({\xi})+{\cdots}\Big\}
  \label{phi-v},
  \end{equation}
  \begin{equation}
 {\phi}_{\phi}^{t}(x) \, =\, 3\,f_{\phi}^{T}\,{\xi}^2
  \label{phi-t},
  \end{equation}
  \begin{equation}
 {\phi}_{\phi}^{s}(x) \, =\, 3\,f_{\phi}^{T}\,{\xi}
  \label{phi-s},
  \end{equation}
 where ${\xi}$ $=$ $x$ $-$ $\bar{x}$; $f_{\phi}$ $=$ $(215{\pm}5)$ MeV
 and $f_{\phi}^{T}$ $=$ $(186{\pm}9)$ MeV \cite{jhep0703.069} are the
 longitudinal and transverse decay constants for the ${\phi}$ meson.
 $C_{2}^{3/2}({\xi})$ is the Gegenbauer polynomial. The nonperturbative
 parameter $a_{2}^{\phi}$ $=$ $0.18{\pm}0.08$ \cite{jhep0703.069} is
 the Gegenbauer moment.

 The $S$-wave two-pion WFs have been defined in Ref.\cite{npb555.231,plb467.263}
  \begin{equation}
 {\Phi}_{{\pi}^{+}{\pi}^{-}}\, =\,
  \frac{1}{4} \Big\{ \!\!\not{q}\,{\phi}_{-}(z,{\zeta},w^{2})
 +{\omega}\,{\phi}_{s}(z,{\zeta},w^{2})
 -{\omega}\,(\!\not{n}_{+}\!\not{n}_{-}-1)\,{\phi}_{+}(z,{\zeta},w^{2}) \Big\}
  \label{wf-2pi-01},
  \end{equation}
 where the variable $z$ gives the momentum fraction of the quark.
 The variables ${\zeta}$ and $w^{2}$ concern the hadronic system but
 not the partons. The asymptotic expressions of the two-pion DAs are
 the variable ${\zeta}$ independent \cite{npb555.231,plb467.263,prd91.094024,epjc76.675},
  \begin{equation}
 {\phi}_{-}(z,{\zeta},w^{2})
  \, =\, 18\,F_{s}(w^{2})\,a_{{\pi}{\pi}}\,z\,\bar{z}\,(\bar{z}-z)
  \, =\, {\phi}_{-}
  \label{da-2pi-m},
  \end{equation}
  \begin{equation}
 {\phi}_{s}(z,{\zeta},w^{2})\, =\, F_{s}(w^{2}) \, =\, {\phi}_{s}
  \label{da-2pi-s},
  \end{equation}
  \begin{equation}
 {\phi}_{+}(z,{\zeta},w^{2})\, =\, F_{s}(w^{2})\,(\bar{z}-z)\, =\, {\phi}_{+}
  \label{da-2pi-p},
  \end{equation}
 where $F_{s}(w^{2})$ is the time-like scalar form factor, and the
 parameter $a_{{\pi}{\pi}}$ $=$ $0.2{\pm}0.2$ \cite{prd91.094024}.
 Clearly, the DAs of ${\phi}_{-}$ and ${\phi}_{+}$ are antisymmetric
 under the interchange $z$ ${\leftrightarrow}$ $\bar{z}$.
 The $F_{s}(w^2)$ involves the strong interaction between the $S$-wave
 resonance and two-pion, as well as elastic rescattering of pion pair.
 Because the mass of the $f_{0}(980)$ meson is near the $K\overline{K}$
 threshold, the form factor $F_{s}(w^{2})$ for the $S$-wave $f_{0}(980)$
 resonance is usually parameterized with the Flatt\'{e} model
 \cite{plb63.228,prd86.052006,prd87.052001,prd89.092006}.
 \begin{equation}
 F_{s}(w^{2})\, =\,
  \frac{m_{f_{0}(980)}^2}
       {m_{f_{0}(980)}^2-w^{2}-im_{f_{0}(980)}
       (g_{{\pi}{\pi}}{\rho}_{{\pi}{\pi}}+
        g_{KK}{\rho}_{KK})}
 \label{ff-f980},
 \end{equation}
 where in our calculation, the mass of the $f_{0}(980)$ meson is fixed to
 the value used by the LHCb Collaboration in the amplitude analysis for
 the $B_{s}$ ${\to}$ ${\phi}\,{\pi}^{+}{\pi}^{-}$ decay, {\em i.e.},
 $m_{f_{0}(980)}$ $=$ 0.98 GeV \cite{prd95.012006}.
 The parameters of $g_{{\pi}{\pi}}$ and $g_{KK}$ are the $f_{0}(980)$
 couplings to the ${\pi}{\pi}$ and $K\overline{K}$ states, respectively.
 Their values are fitted by the LHCb Collaboration through the $B_{s}$ ${\to}$
 $J/{\psi}{\pi}^{+}{\pi}^{-}$ decay, and $g_{{\pi}{\pi}}$ $=$ $167{\pm}7$ MeV
 and $g_{KK}$ $=$ $(3.47{\pm}0.12)g_{{\pi}{\pi}}$ \cite{prd89.092006}.
 The expressions of the phase space factors are written as
 \cite{prd86.052006,prd87.052001,prd89.092006}:
  \begin{equation}
 {\rho}_{{\pi}{\pi}}\, =\,
  \frac{2}{3}\,\sqrt{ 1-\frac{ 4m^{2}_{{\pi}^{\pm}} }{ w^{2} } }
 +\frac{1}{3}\,\sqrt{ 1-\frac{ 4m^{2}_{{\pi}^{0}} }{ w^{2} } }
  \label{rho-pipi},
  \end{equation}
  \begin{equation}
 {\rho}_{KK}\, =\,
  \frac{1}{2}\,\sqrt{ 1-\frac{ 4m^{2}_{K^{\pm}} }{ w^{2} } }
 +\frac{1}{2}\,\sqrt{ 1-\frac{ 4m^{2}_{K^{0}} }{ w^{2} } }
  \label{rho-kk}.
  \end{equation}

  \subsection{Deacy amplitude}
  \label{sec0205}
  The Feynman diagrams for the $B_{s}$ ${\to}$ ${\phi}\,f_{0}(980)$ ${\to}$
  ${\phi}\,\pi^{+}\pi^{-}$ decay within the pQCD framework are show in
  Fig.\ref{fey}, including (1) the ${\phi}$ meson emission while the $B_{s}$
  meson transition into the two-pion pair through the $f_{0}(980)$ resonance
  in Fig.\ref{fey}(a-d), (2) the two-pion pair emission while the $B_{s}$ meson
  transition into the ${\phi}$ meson in Fig.\ref{fey}(e-h), (3) the $B_{s}$
  annihilation in Fig.\ref{fey}(i-p). In addition, the diagrams in the first
  (last) two columns are called the (non)factorizable topologies.
  In general, the amplitudes of the factorizable topologies have the relatively
  simple structures. For the factorizable topologies of Fig.\ref{fey}(a,b),
  the ${\phi}$ meson can be isolated from the $B_{s}{\pi}{\pi}$ system,
  so the amplitudes can be written as the product of the decay constant $f_{\phi}$
  and the $B_{s}$ ${\to}$ ${\pi}{\pi}$ transition form factors.
  Similarly, for the factorizable topologies of Fig.\ref{fey}(e,f),
  the two-pion pair can be isolated from the $B_{s}{\phi}$ system, and the
  transition matrix elements between the vacuum and the two-pion pair can
  be expressed as the time-like scalar form factor $F_{s}(w^{2})$.
  So the HMEs of the local operators can be written as the $B_{s}$ ${\to}$
  ${\phi}$ transition form factors multiplied by the form factor $F_{s}(w^{2})$.
  Likewise, for the factorizable topologies of Fig.\ref{fey}(i,j) and (m,n),
  the $B_{s}$ meson can be isolated from the final states, so the amplitudes
  can be written as the product of the decay constant $f_{B_{s}}$ and
  the time-like form factors for the transition between the ${\phi}$ meson
  and the two-pion pair.
  The amplitudes of the nonfactorizable topologies involve the DAs of all
  participating mesons.

  Using the PQCD formula in Eq.(\ref{hme-quasi}) for the quasi two-body decay,
  the amplitude for the $B_{s}$ ${\to}$ ${\phi}\,f_{0}(980)$ ${\to}$
  ${\phi}\,\pi^{+}\pi^{-}$ decay is written as follow.
  \begin{eqnarray}
 {\cal A} &=& \, \Big(V_{ub}^{\ast}\,V_{us}+V_{cb}^{\ast}\,V_{cs}\Big)\,
  \Big\{ {\cal A}_{ef}^{LL}[a_{3} - \frac{1}{2} a_{9}]
       + {\cal A}_{ef}^{LR}[a_{5} - \frac{1}{2} a_{7}]
       + {\cal A}_{ef}^{SP}[a_{6} - \frac{1}{2} a_{8}]
  \nonumber \\ & & \quad
       + {\cal A}_{nf}^{LL}[C_{4} - \frac{1}{2} C_{10}]
       + {\cal A}_{nf}^{LR}[C_{6} - \frac{1}{2} C_{8} ]
       + {\cal A}_{nf}^{SP}[C_{5} - \frac{1}{2} C_{7} ] \Big\}
  \label{amp},
  \end{eqnarray}
  \begin{equation}
 {\cal A}_{ef}^{\rho}\, =\, {\cal A}_{a}^{\rho}+{\cal A}_{b}^{\rho}
                          + {\cal A}_{e}^{\rho}+{\cal A}_{f}^{\rho}
                          + {\cal A}_{i}^{\rho}+{\cal A}_{j}^{\rho}
                          + {\cal A}_{m}^{\rho}+{\cal A}_{n}^{\rho},
  \quad \text{for } {\rho}\,=\, LL,LP,SP
  \label{a-ef}
  \end{equation}
  \begin{equation}
 {\cal A}_{nf}^{\rho}\, =\, {\cal A}_{c}^{\rho}+{\cal A}_{d}^{\rho}
                          + {\cal A}_{g}^{\rho}+{\cal A}_{h}^{\rho}
                          + {\cal A}_{k}^{\rho}+{\cal A}_{l}^{\rho}
                          + {\cal A}_{o}^{\rho}+{\cal A}_{p}^{\rho},
  \quad \text{for } {\rho}\,=\, LL,LP,SP
  \label{a-nf}
  \end{equation}
  \begin{equation}
   a_{i}\, =\, \Bigg\{ \begin{array}{l}
   C_{i}+C_{i+1}/N_{c}  \quad \text{for odd }i; \\
   C_{i}+C_{i-1}/N_{c}  \quad \text{for even }i,
  \end{array}
  \label{eq:ai}
  \end{equation}
  where the Wilson coefficients $C_{i}$ are looked as the function variables
  of the amplitudes of ${\cal A}_{ef}^{\rho}$, ${\cal A}_{nf}^{\rho}$ and
  ${\cal A}_{\sigma}^{\rho}$, and $N_{c}$ $=$ $3$ is the color number.
  ${\cal A}_{ef}^{\rho}$ (${\cal A}_{ff}^{\rho}$) is the sum of the amplitudes
  for the (non)factorizable topologies. The superscript ${\rho}$ of the
  amplitude building block ${\cal A}_{\sigma}^{\rho}$ refers to the three
  possible Dirac structures ${\Gamma}_{1}{\otimes}{\Gamma}_{2}$ of the operators
  $(\bar{q}_{1}q_{2})_{{\Gamma}_{1}}(\bar{q}_{3}q_{4})_{{\Gamma}_{2}}$,
  namely ${\rho}$ $=$ $LL$ for $(V-A){\otimes}(V-A)$, ${\rho}$ $=$ $LR$
  for $(V-A){\otimes}(V+A)$ and ${\rho}$ $=$ $SP$ for $-2(S-P){\otimes}(S+P)$.
  The subscript ${\sigma}$ of ${\cal A}_{\sigma}^{\rho}$ (${\sigma}$ $=$ $a$
  $b$, ${\cdots}$, $p$) corresponds to the sub-diagram indices of Fig.\ref{fey}.
  ${\cal A}_{ef}^{\rho}$, ${\cal A}_{ff}^{\rho}$ and ${\cal A}_{\sigma}^{\rho}$
  are the functions of the Wilson coefficient $C_{i}$.
  The analytical expressions of the amplitude building blocks ${\cal A}_{\sigma}^{\rho}$
  are listed in Appendix \ref{block} in detail.

  \section{numerical results and discussion}
  \label{sec03}
   The differential branching ratio for the sequential $B_{s}$ ${\to}$
   ${\phi}\,f_{0}(980)$ ${\to}$ ${\phi}\,\pi^{+}\pi^{-}$ decay is \cite{pdg}:
  \begin{equation}
  \frac{d{\cal B}}{d w}\, =\,
  \frac{ {\tau}_{B_{s}}\,p_{\pi}^{\ast}\,p_{\rm cm}}
       {4\,(2{\pi})^{3}\, m_{B_{s}}^{2}}
       {\vert}{\cal A}{\vert}^{2}
  \label{eq:br},
  \end{equation}
  where ${\tau}_{B_{s}}$ $=$ $(1.510{\pm}0.005)$ ps is the lifetime of
  the $B_{s}$ meson \cite{pdg}. The kinematic variable $p_{\pi}^{\ast}$
  is the pion momentum in the rest frame of the two-pion pair,
  \begin{equation}
  p_{\pi}^{\ast}\, =\, \frac{1}{2}\,\sqrt{w^{2}-4m_{{\pi}^{\pm}}^{2}}
  \label{kine-p-pi}.
  \end{equation}

  In our calculation, besides the aforementioned parameters, other related
  parameters, such as the mass of the mesons and quarks, will take their
  values given in Ref.\cite{pdg}. And if it is not specified explicitly,
  their central values will be fixed as the default inputs.
  Our numerical result of the branching ratio is
  \begin{equation}
  {\cal B}(B_{s}{\to}{\phi}\,f_{0}(980){\to}{\phi}\,\pi^{+}\pi^{-})\, =\,
   \Big[1.31_{-0.31}^{+0.40}(a_{{\pi}{\pi}}){}_{-0.16}^{+0.19}(m_{b})
   {}_{-0.09}^{+0.10}(\text{CKM})\Big]{\times}10^{-6}
  \label{num-br},
  \end{equation}
  where the uncertainties come from the parameter $a_{{\pi}{\pi}}$ of
  DA in Eq.(\ref{da-2pi-m}), the $b$ quark mass $m_{b}$, the CKM factors
  $V_{ub}^{\ast}V_{us}$ and $V_{cb}^{\ast}V_{cs}$, respectively.
  It is clear that the result in Eq.(\ref{num-br}) agrees with the LHCb
  measurement in Eq.(\ref{eq:br-exp-01}) within uncertainties.

  \begin{figure}[h]
  \includegraphics[width=0.7\textwidth,bb=15 50 540 530]{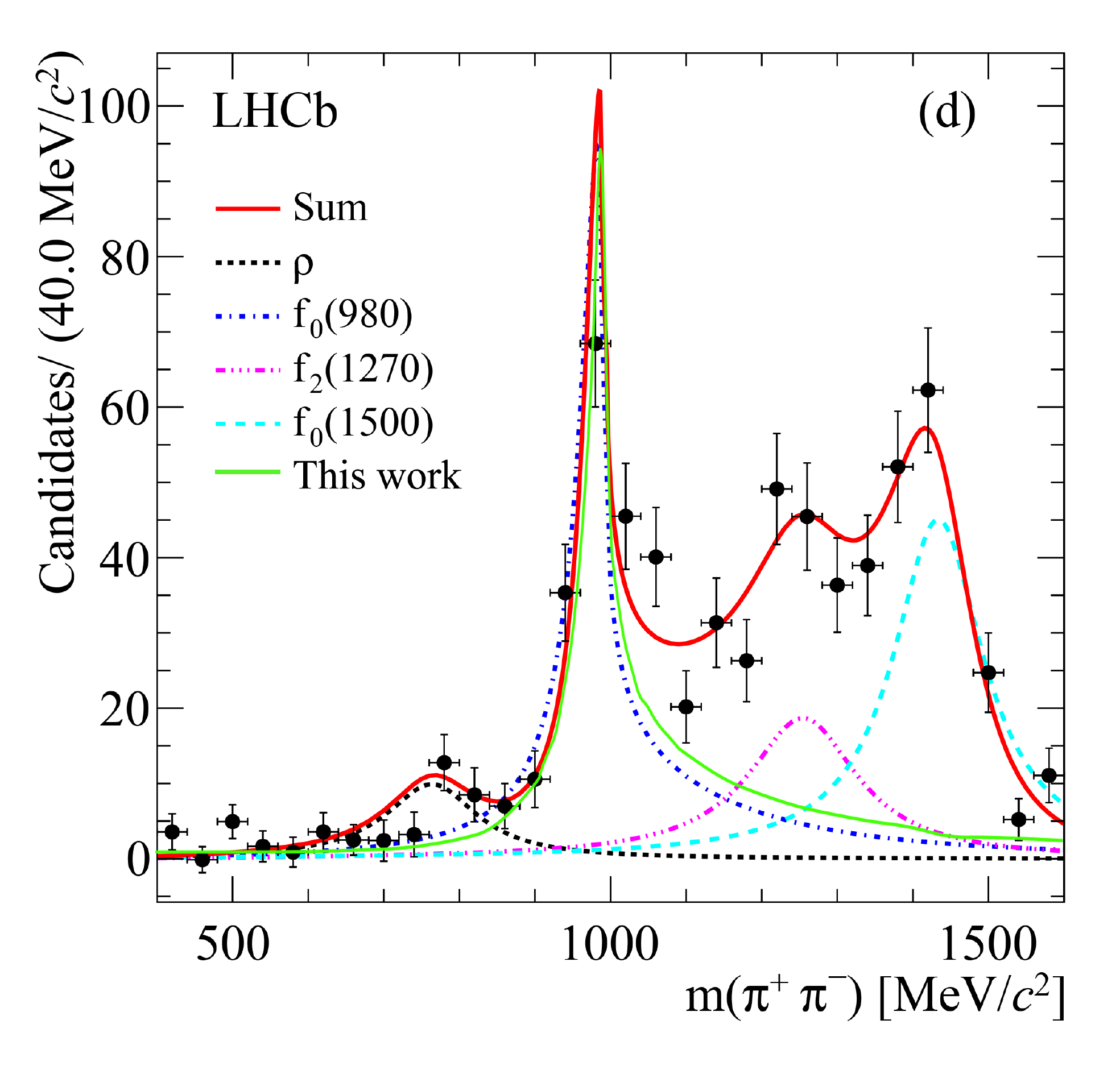}
  \caption{The distributions from some resonances versus the invariant mass
  of the ${\pi}^{+}{\pi}^{-}$ pair for the $B_{s}$ ${\to}$ ${\phi}\,f_{0}(980)$
  ${\to}$ ${\phi}\,\pi^{+}\pi^{-}$ decay, where the solid (green) line
  is our result with the PQCD approach, the dot-dashed (blue) line is
  the $f_{0}(980)$ meson contribution given by the LHCb Collaboration
  in Fig.7(d) of Ref.\cite{prd95.012006}, and a full explanation of
  other lines can be found in Ref.\cite{prd95.012006}.}
  \label{fig:f980}
  \end{figure}

  To illustrate the $S$-wave $f_{0}(980)$ contribution to the decay in question,
  and to compare our result with the experimental measurement, the dependence
  of the calibrated differential branching ratio $d{\cal B}/dw$ on the pion-pair
  invariant mass $w$ $=$ $m({\pi}^{+}{\pi}^{-})$ is shown in Fig.\ref{fig:f980}.
  It is seen that the result with the PQCD approach is generally consistent
  with the shape line of the $f_{0}(980)$ meson fitted by the LHCb Collaboration
  \cite{prd95.012006}.

  In addition, the contributions from different topologies to the $B_{s}$ ${\to}$
  ${\phi}\,f_{0}(980)$ ${\to}$ ${\phi}\,\pi^{+}\pi^{-}$ decay are investigated.
  It shows that
  (1) the main contributions come from the factorizable ${\phi}$
  emission topologies of Fig.\ref{fey}(a,b).
  (2) Due to the renormalization conditions of the two-pion DAs, only the
  amplitudes corresponding to the Dirac current structure of
  ${\Gamma}_{1}{\otimes}{\Gamma}_{2}$ $=$ $-2(S-P){\otimes}(S+P)$ have
  nonzero contributions [see Eq.(\ref{amp-e-lr})-Eq.(\ref{amp-f-sp})] for
  the factorizable two-pion emission topologies of Fig.\ref{fey}(e,f).
  (3) Because of the opposite sign of the quark propagators between the
  factorizable annihilation topologies of Fig.\ref{fey}(i) [Fig.\ref{fey}(j)]
  and Fig.\ref{fey}(n) [Fig.\ref{fey}(m)], the interference cancelation
  mechanism results in the relatively small total contributions from the
  factorizable annihilation topologies.
  (4) For each type diagrams, such as the ${\phi}$ emission diagrams
  in Fig.\ref{fey}(a-d) or the two-pion emission diagrams in Fig.\ref{fey}(e-h),
  the nonfactorizable contributions are small relative to the factorizable
  contributions because of the $1/N_{c}$ suppression.
  (5) The relative magnitudes of decay amplitudes basically correspond with
  the power estimations in Ref.\cite{plb561.258}, {\em i.e.},
  \begin{equation}
   \sum\limits_{i=a,b,c,d \atop {\alpha}=LL,LR,SP}\!\!\!\!\!\! {\cal A}_{i}^{\alpha}
   \, :\!\!\!\!\!\!\!
   \sum\limits_{j=e,f,g,h \atop {\beta}=LL,LR,SP}\!\!\!\!\!\! {\cal A}_{j}^{\beta}
   \, :\!\!\!\!\!\!\!
   \sum\limits_{k=i,{\cdots},p \atop {\rho}=LL,LR,SP}\!\!\!\!\!\! {\cal A}_{k}^{\rho}
   \, =\, 1\, :\, \frac{w}{m_{B_{s}}}\, :\, \frac{{\Lambda}_{\rm QCD}}{m_{B_{s}}}
  \label{eq:power}.
  \end{equation}

  It should be pointed out that the $B_{s}$ ${\to}$ ${\phi}\,\pi^{+}\pi^{-}$ decay
  could be approximately handled as the quasi two-body sequential $B_{s}$ ${\to}$
  ${\phi}\,f_{0}(980)$ ${\to}$ ${\phi}\,\pi^{+}\pi^{-}$ decay at the edges
  of the Dalitz plot, and there are still many factors that can affect the
  theoretical result. For example, the contributions from the center regions of
  the Dalitz plot and the nonresonant contributions to the $B_{s}$ ${\to}$
  ${\phi}\,\pi^{+}\pi^{-}$ decay are not considered in this paper.
  It has shown in Refs.\cite{prd88.114014,prd89.094007,prd94.094015} that the
  nonresonant contributions are important and deserve much attention, which is
  beyond the scope of this paper.

  \section{summary}
  \label{sec04}
  The rare cascade $B_{s}$ ${\to}$ ${\phi}f_{0}(980)$ ${\to}$
  ${\phi}\,{\pi}^{+}{\pi}^{-}$ decay is induced by the FCNC
  $b$ ${\to}$ $\bar{s}s\bar{s}$ process within SM, where the
  isoscalar $f_{0}(980)$ meson has a substantial $s\bar{s}$
  component. Given the two-pion pair with small invariant mass
  GeV comes from the $S$-wave resonant $f_{0}(980)$ state,
  the three-body $B_{s}$ ${\to}$ ${\phi}f_{0}(980)$ ${\to}$
  ${\phi}\,{\pi}^{+}{\pi}^{-}$ decay can be approximated
  as the quasi two-body decay.
  By introducing the nonperturbative two-pion DAs to describe
  the two-pion system, and parameterizing the scalar form factor
  for the $f_{0}(980)$ resonance with the Flatt\'{e} model, the
  $B_{s}$ ${\to}$ ${\phi}f_{0}(980)$ ${\to}$
  ${\phi}\,{\pi}^{+}{\pi}^{-}$ decay is studied with the PQCD
  approach. It is found that with appropriate parameters, the
  theoretical result of the branching ratio in the mass range
  400 MeV $<$ $m({\pi}^{+}{\pi}^{-})$ $<$ 1600 MeV is in agreement
  with the recent LHCb data \cite{prd95.012006} within uncertainties.

  \section*{Acknowledgments}
  The work is supported by the National Natural Science Foundation
  of China (Grant Nos. 11705047, U1632109, 11547014 and 11475055).

  \begin{appendix}
  \section{The amplitude building blocks for the $B_{s}$ ${\to}$
  ${\phi}\,f_{0}(980)$ ${\to}$ ${\phi}\,\pi^{+}\pi^{-}$ decay}
  \label{block}
  \begin{equation}
 {\cal F}\, =\, -i\,{\pi}\,C_{F}\,f_{B_{s}}
  \label{amp-f},
  \end{equation}
  \begin{eqnarray}
 {\cal A}_{a}^{LL}[C_{i}] &=& 2\, {\cal F}\, f_{\phi}\,
 {\int}^{1}_{0} dx\,dz {\int}_{0}^{\infty} db\,db_{f}\, b\,b_{f}\,
 {\alpha}_{s}(t_{a})\, H_{ef}({\alpha}_{a},{\beta}_{a},b,b_{f})\,
 E_{\phi}(t_{a})\,C_{i}(t_{a})
  \nonumber \\ &{\times}&  m_{B_{s}}
  \Big\{ {\phi}_{B}^{a}(x)\, m_{B_{s}}
  \Big[ {\phi}_{-}\, m_{B_{s}} p_{\rm cm}\, (1-z+z\,r_{\phi}^{2})
      + {\phi}_{s}\, r_{b}\, w\, p_{\rm cm}
  \nonumber \\ &+&
        {\phi}_{+}\, r_{b}\, w\, E_{\phi}  \Big]
  - 2\, {\phi}_{B}^{p}(x)\, \Big[ {\phi}_{+}\, w\,
  ( m_{B_{s}} E_{\phi} + z\, w^{2} - z\, E_{w}\, m_{B_{s}} )
  \nonumber \\ &+&
        {\phi}_{-}\, r_{b}\, m_{B_{s}}^{2} p_{\rm cm}
      + {\phi}_{s}\, m_{B_{s}} w\, p_{\rm cm}\, (1-z)
  \Big] \Big\}
  \label{amp-a-ll},
  \end{eqnarray}
  \begin{equation}
 {\cal A}_{a}^{LR}[C_{i}]\, =\, {\cal A}_{a}^{LL}[C_{i}]
  \label{amp-a-lr},
  \end{equation}
  \begin{equation}
 {\cal A}_{a}^{SP}[C_{i}]\, =\, 0
  \label{amp-a-sp},
  \end{equation}
  \begin{eqnarray}
 {\cal A}_{b}^{LL}[C_{i}] &=& 2\, {\cal F}\, f_{\phi}\,
 {\int}^{1}_{0} dx\,dz {\int}_{0}^{\infty} db\,db_{f}\, b\,b_{f}\,
 {\alpha}_{s}(t_{b})\, H_{ef}({\alpha}_{a},{\beta}_{b},b_{f},b)\,
  E_{\phi}(t_{b})\,C_{i}(t_{b})
  \nonumber \\ &{\times}& m_{B_{s}}  p_{\rm cm}\,
  \Big\{ {\phi}_{B}^{a}(x)\, {\phi}_{-}\, ( \bar{x}\,w^2+x\,m_{\phi}^2 )
      -  {\phi}_{B}^{p}(x)\,{\phi}_{s}\,2\,m_{B_{s}} w\, \bar{x} \Big\}
  \label{amp-b-ll},
  \end{eqnarray}
  \begin{equation}
 {\cal A}_{b}^{LR}[C_{i}]\, =\, {\cal A}_{b}^{LL}[C_{i}]
  \label{amp-b-lr},
  \end{equation}
  \begin{equation}
 {\cal A}_{b}^{SP}[C_{i}]\, =\, 0
  \label{amp-b-sp},
  \end{equation}
  \begin{eqnarray}
 {\cal A}_{c}^{LL}[C_{i}] &=&
  \frac{{\cal F}}{N_{c}} {\int}^{1}_{0} dx\,dy\,dz
 {\int}_{0}^{\infty} db\, db_{\phi}\, db_{f}\, b_{\phi}\, b_{f}\,
 {\alpha}_{s}(t_{c})\, H_{en}({\alpha}_{a},{\beta}_{c},b_{\phi},b,b_{f})\,
 E_{n}(t_{c})\,C_{i}(t_{c})
  \nonumber \\ &{\times}& {\phi}_{\phi}^{v}(y)\,
  \Big\{ 4\,m_{B_{s}}^{2} p_{\rm cm}\,{\phi}_{B}^{a}(x)\,
 {\phi}_{-}\, (y\,E_{\phi}+z\,E_{w}-x\,m_{B_{s}})
 +\Big[ {\phi}_{s}\,m_{B_{s}} p_{\rm cm}\,(x-z)
  \nonumber \\ &+&
  {\phi}_{+}\, (x\,m_{B_{s}} E_{\phi}-y\,m_{\phi}^{2}-z\,m_{B_{s}} E_{w}+z\,w^2)
  \Big] 2\,m_{B_{s}}  w\, {\phi}_{B}^{p}(x)\, \Big\}
  \label{amp-c-ll},
  \end{eqnarray}
  \begin{eqnarray}
 {\cal A}_{c}^{LR}[C_{i}] &=&
  \frac{{\cal F}}{N_{c}} {\int}^{1}_{0} dx\,dy\,dz
 {\int}_{0}^{\infty} db\, db_{\phi}\, db_{f}\, b_{\phi}\, b_{f}\,
 {\alpha}_{s}(t_{c})\, H_{en}({\alpha}_{a},{\beta}_{c},b_{\phi},b,b_{f})\,
 E_{n}(t_{c})\,C_{i}(t_{c})
  \nonumber \\ &{\times}& 2\,m_{B_{s}} {\phi}_{\phi}^{v}(y)\,
  \Big\{ 2\,p_{\rm cm}\,{\phi}_{B}^{a}(x)\,
 {\phi}_{-}\, (y\,p_{\rm cm}^2+y\,E_{w}\,E_{\phi}
 +z\,w^2-x\,E_{w}\,m_{B_{s}})
  \nonumber \\ &+& w\, {\phi}_{B}^{p}(x)
  \Big[ {\phi}_{s}\,m_{B_{s}} p_{\rm cm}\,(x-z)
  - {\phi}_{+}\, (x\,m_{B_{s}} E_{\phi}-y\,m_{\phi}^{2}
  - z\,m_{B_{s}} E_{w}+z\,w^2) \Big] \Big\},\qquad
  \label{amp-c-lr}
  \end{eqnarray}
  \begin{eqnarray}
 {\cal A}_{c}^{SP}[C_{i}] &=&
  \frac{-2\,{\cal F}}{N_{c}} {\int}^{1}_{0} dx\,dy\,dz
 {\int}_{0}^{\infty} db\, db_{\phi}\, db_{f}\, b_{\phi}\, b_{f}\,
 {\alpha}_{s}(t_{c})\, H_{en}({\alpha}_{a},{\beta}_{c},b_{\phi},b,b_{f})\,
 E_{n}(t_{c})\,C_{i}(t_{c})
  \nonumber \\ &{\times}&
  m_{B_{s}}  m_{\phi}\, \Big\{ {\phi}_{\phi}^{t}(y)\,
  \Big[ {\phi}_{B}^{a}(x)\,w\, \{ {\phi}_{s}\,p_{\rm cm}\,(y-z)
  - {\phi}_{+}\,(y\,E_{\phi}+z\,E_{w}-x\,m_{B_{s}}) \}
  \nonumber \\ &-&
 {\phi}_{B}^{p}(x)\,{\phi}_{-}\,m_{B_{s}} p_{\rm cm}\,(x-y)\Big]
  + {\phi}_{\phi}^{s}(y)\,\Big[ {\phi}_{B}^{a}(x)\,w\, \{
 {\phi}_{s}\,(y\,E_{\phi}+z\,E_{w}-x\,m_{B_{s}})
  \nonumber \\ &+&
 {\phi}_{+}\,p_{\rm cm}\,(z-y) \} - {\phi}_{B}^{p}(x)\,{\phi}_{-}\,
 (x\,m_{B_{s}} E_{w}-y\,E_{\phi}\,E_{w}-y\,p_{\rm cm}^2 -z\,w^2)
  \Big] \Big\}
  \label{amp-c-sp},
  \end{eqnarray}
  \begin{eqnarray}
 {\cal A}_{d}^{LL}[C_{i}] &=&
  \frac{{\cal F}}{N_{c}} {\int}^{1}_{0} dx\,dy\,dz
 {\int}_{0}^{\infty} db\, db_{\phi}\, db_{f}\, b_{\phi}\, b_{f}\,
 {\alpha}_{s}(t_{d})\, H_{en}({\alpha}_{a},{\beta}_{d},b_{\phi},b,b_{f})\,
 E_{n}(t_{d})\,C_{i}(t_{d})
  \nonumber \\ &{\times}& 2\,m_{B_{s}}\, {\phi}_{\phi}^{v}(y)\,
  \Big\{ {\phi}_{B}^{a}(x)\,{\phi}_{-}\,2\, p_{\rm cm}\,
  \Big[ x\,m_{B_{s}} E_{w}-\bar{y}\,(m_{B_{s}} E_{\phi}-m_{\phi}^2) -z\,w^2 \Big]
  \nonumber \\ &+& {\phi}_{B}^{p}(x)\,w
  \Big[ {\phi}_{s}\,m_{B_{s}} p_{\rm cm}\, (z-x)
     + {\phi}_{+}\, (x\,m_{B_{s}} E_{\phi} -\bar{y}\,m_{\phi}^{2}
     - z\,m_{B_{s}} E_{w}+z\,w^2 ) \Big] \Big\},\qquad
  \label{amp-d-ll}
  \end{eqnarray}
  \begin{eqnarray}
 {\cal A}_{d}^{LR}[C_{i}] &=&
  \frac{{\cal F}}{N_{c}} {\int}^{1}_{0} dx\,dy\,dz
 {\int}_{0}^{\infty} db\, db_{\phi}\, db_{f}\, b_{\phi}\, b_{f}\,
 {\alpha}_{s}(t_{d})\, H_{en}({\alpha}_{a},{\beta}_{d},b_{\phi},b,b_{f})\,
 E_{n}(t_{d})\,C_{i}(t_{d})
  \nonumber \\ &{\times}& 2\,m_{B_{s}}\, {\phi}_{\phi}^{v}(y)\, \Big\{
 {\phi}_{B}^{a}(x)\,{\phi}_{-}\,2\, m_{B_{s}} p_{\rm cm}\,
  \Big[ x\,m_{B_{s}}-\bar{y}\,E_{\phi}-z\,E_{w} \Big]
  \nonumber \\ &+& {\phi}_{B}^{p}(x)\,w
  \Big[ {\phi}_{s}\,m_{B_{s}} p_{\rm cm}\,(z-x)
      - {\phi}_{+}\,(x\,m_{B_{s}}E_{\phi}-\bar{y}\,m_{\phi}^{2}
      - z\,m_{B_{s}} E_{w}+z\,w^2 ) \Big] \Big\},\qquad
  \label{amp-d-lr}
  \end{eqnarray}
  \begin{eqnarray}
 {\cal A}_{d}^{SP}[C_{i}] &=&
  \frac{-2\,{\cal F}}{N_{c}} {\int}^{1}_{0} dx\,dy\,dz
 {\int}_{0}^{\infty} db\, db_{\phi}\, db_{f}\, b_{\phi}\, b_{f}\,
 {\alpha}_{s}(t_{d})\, H_{en}({\alpha}_{a},{\beta}_{d},b_{\phi},b,b_{f})\,
 E_{n}(t_{d})\,C_{i}(t_{d})
  \nonumber \\ &{\times}& m_{B_{s}}  m_{\phi}\, \Big\{
 {\phi}_{\phi}^{t}(y)\,\Big[ {\phi}_{B}^{a}(x)\,w\,
  \{ {\phi}_{s}\,p_{\rm cm}\,(\bar{y}-z)
  +  {\phi}_{+}\,( y\,E_{w}-E_{\phi}+x\,m_{B_{s}}-z\,E_{w} ) \}
  \nonumber \\ &-&
 {\phi}_{B}^{p}(x)\,{\phi}_{-}\,m_{B_{s}} p_{\rm cm}\,(x-\bar{y}) \Big]
  + {\phi}_{\phi}^{s}(y)\,\Big[ {\phi}_{B}^{a}(x)\,w\, \{ {\phi}_{s}\,
  (x\,m_{B_{s}}-z\,E_{w}+y\,E_{\phi}-E_{\phi})
  \nonumber \\ &+&
  {\phi}_{+}\,p_{\rm cm}(\bar{y}-z) \}
 +{\phi}_{B}^{p}(x)\,{\phi}_{-}\,( x\,m_{B_{s}}E_{w}
 -\bar{y}\,E_{\phi}\,E_{w} -\bar{y}\,p_{\rm cm}^2-z\,w^2)
  \Big] \Big\}
  \label{amp-d-sp},
  \end{eqnarray}
  \begin{equation}
 {\cal A}_{e}^{LL}[C_{i}]\, =\, {\cal A}_{e}^{LR}[C_{i}]\, =\, 0
  \label{amp-e-lr},
  \end{equation}
  \begin{eqnarray}
 {\cal A}_{e}^{SP}[C_{i}] &=& -4\, m_{B_{s}}\, {\cal F}\,
 {\int}^{1}_{0} dx\,dy {\int}_{0}^{\infty} db\, db_{\phi}\, b\, b_{\phi}\,
 {\alpha}_{s}(t_{e})\, H_{ef}({\alpha}_{e},{\beta}_{e},b,b_{\phi})\,
 E_{f}(t_{e})\,C_{i}(t_{e})
  \nonumber \\ &{\times}& {\phi}_{s}\,w\,\Big\{ {\phi}_{B}^{a}(x)
  \Big[ {\phi}_{\phi}^{s}(y)\, m_{\phi}\, (m_{B_{s}}-y\,E_{\phi})
  - {\phi}_{\phi}^{v}(y)\,m_{b}\,p_{\rm cm}
  - {\phi}_{\phi}^{t}(y)\,m_{\phi}\, p_{\rm cm}\,y \Big]
  \nonumber \\ &+& 2\,
 {\phi}_{B}^{p}(x)\, \Big[ {\phi}_{\phi}^{v}(y)\, m_{B_{s}} p_{\rm cm}
  -{\phi}_{\phi}^{s}(y)\,m_{\phi}\,m_{b}\ \Big] \Big\}
  \label{amp-e-sp},
  \end{eqnarray}
  \begin{equation}
 {\cal A}_{f}^{LL}[C_{i}]\, =\, {\cal A}_{f}^{LR}[C_{i}]\, =\, 0
  \label{amp-f-lr},
  \end{equation}
  \begin{eqnarray}
 {\cal A}_{f}^{SP}[C_{i}] &=& -4\, m_{B_{s}} {\cal F}\,
 {\int}^{1}_{0} dx\,dy {\int}_{0}^{\infty} db\, db_{\phi}\, b\, b_{\phi}\,
 {\alpha}_{s}(t_{f})\, H_{ef}({\alpha}_{e},{\beta}_{f},b_{\phi},b)\,
  E_{f}(t_{f})\,C_{i}(t_{f})
  \nonumber \\ &{\times}& {\phi}_{s}\,w\,\Big\{
 {\phi}_{B}^{a}(x)\,{\phi}_{\phi}^{s}(y)\,2\,m_{\phi}\,(E_{\phi}-x\,m_{B_{s}})
-{\phi}_{B}^{p}(x)\,{\phi}_{\phi}^{v}(y)\, m_{B_{s}} p_{\rm cm}\,x \Big\}
  \label{amp-f-sp},
  \end{eqnarray}
\begin{eqnarray}
{\cal A}_{g}^{LL}[C_{i}] &=& \frac{{\cal F}}{N_{C}} {\int}^{1}_{0} dx\,dy\,dz {\int}_{0}^{\infty} b db  \, b_{\phi}db_{\phi}\, b_{f} db_{f}\, {\alpha}_{s}(t_{g})\, H_{en}({\alpha}_{e},\, {\beta}_{g},\, b_{f},\, b_{\phi},\, b)\, E_{n}(t_{g})\,C_{i}(t_{g})\nonumber\\
&& \times \,
2\,m_{B_{s}}^2\,{\phi}_{-}\,\Big\{{\phi}_{B}^{a}(x)\,{\phi}_{\phi}^{v}(y)\,2\,p_{\rm cm}
\,(y\,E_{\phi}+z\,E_{w}-x\,m_{B_{s}})\nonumber\\
&&+\,{\phi}_{B}^{p}(x)\,r\,\bigg[{\phi}_{\phi}^{s}(y)\,
\big[E_{w}\,(y\,E_{\phi}-x\,m_{B_{s}})+y\,p_{\rm cm}^2+z\,w^2 \big]\nonumber\\
&&
 -\,{\phi}_{\phi}^{t}(y)\,m_{B_{s}}\,p_{\rm cm}(y-x)
\bigg]\Big\}
\label{amp-g-ll},
\end{eqnarray}
\begin{eqnarray}
{\cal A}_{g}^{LR}[C_{i}] &=& \frac{{\cal F}}{N_{C}} {\int}^{1}_{0} dx\,dy\,dz {\int}_{0}^{\infty} b db  \, b_{\phi}db_{\phi}\, b_{f} db_{f}\, {\alpha}_{s}(t_{g})\, H_{en}({\alpha}_{e},\, {\beta}_{g},\, b_{f},\, b_{\phi},\, b)\, E_{n}(t_{g})\,C_{i}(t_{g})\nonumber\\
&& \times \,
2\,m_{B_{s}}^2\,{\phi}_{-}\,\Big\{{\phi}_{B}^{a}(x)\,{\phi}_{\phi}^{v}(y)\,2\,E_{w}\,p_{\rm cm}
\,(z-x)\nonumber\\
&&+\,{\phi}_{B}^{p}(x)\,r\,\bigg[{\phi}_{\phi}^{s}(y)\,
\big[E_{w}\,(y\,E_{\phi}-x\,m_{B_{s}})+y\,p_{\rm cm}^2+z\,w^2\big]\nonumber\\
&&+\,{\phi}_{\phi}^{t}(y)\,m_{B_{s}}\,p_{\rm cm}(y-x)\bigg]\Big\}
\label{amp-g-lr},
\end{eqnarray}
\begin{eqnarray}
{\cal A}_{g}^{SP}[C_{i}] &=& \frac{(-2){\cal F}}{N_{C}} {\int}^{1}_{0} dx\,dy\,dz {\int}_{0}^{\infty} b db  \, b_{\phi}db_{\phi}\, b_{f} db_{f}\, {\alpha}_{s}(t_{g})\, H_{en}({\alpha}_{e},\, {\beta}_{g},\, b_{f},\, b_{\phi},\, b)\, E_{n}(t_{g})\,C_{i}(t_{g})\nonumber\\
&& \times \,m_{B_{s}}\,w\,
\Big\{r\,m_{B_{s}}\,{\phi}_{B}^{a}(x)\,\bigg[({\phi}_{\phi}^{t}(y)\,{\phi}_{s}\,-
{\phi}_{\phi}^{s}(y)\,{\phi}_{+})\,p_{\rm cm}\,
(y-z)\nonumber\\
&&+ \,({\phi}_{\phi}^{t}(y)\,{\phi}_{+}\,-
{\phi}_{\phi}^{s}(y)\,{\phi}_{s}) (y\,E_{\phi}-x\,m_{B_{s}}+z\,E_{w})\bigg]\nonumber\\
&&+\,{\phi}_{B}^{p}(x)\,{\phi}_{\phi}^{v}(y)\,\Big[{\phi}_{s}\,m_{B_{s}}\,p_{\rm cm}\,(z-x)\nonumber\\
&&+\,{\phi}_{+}\,(x\,m_{B_{s}}\,E_{\phi}-y\,m_{\phi}^2+z\,w^2
-z\,E_{w}\,m_{B_{s}})\Big]\Big\}
\label{amp-g-sp},
\end{eqnarray}
\begin{eqnarray}
{\cal A}_{h}^{LL}[C_{i}] &=& \frac{{\cal F}}{N_{C}} {\int}^{1}_{0} dx\,dy\,dz {\int}_{0}^{\infty} b db  \, b_{\phi}db_{\phi}\, b_{f} db_{f}\, {\alpha}_{s}(t_{h})\, H_{en}({\alpha}_{e},\, {\beta}_{h},\, b_{f},\, b_{\phi},\, b)\, E_{n}(t_{h})\,C_{i}(t_{h})\nonumber\\
&& \times \,2\,m_{B_{s}}^2\,
{\phi}_{-}\,\Big\{{\phi}_{B}^{a}(x)\,{\phi}_{\phi}^{v}(y)\,2\,E_{w}\,p_{\rm cm}
\,(x+z-1)\nonumber\\
&& +\,{\phi}_{B}^{p}(x)\,r\,\bigg[{\phi}_{\phi}^{s}(y)
\big[E_{w}(x\,m_{B_{s}}-y\,E_{\phi})-y\,p_{\rm cm}^2+(z-1)\,w^2\big]\nonumber\\
&&
 -\,{\phi}_{\phi}^{t}(y)\,m_{B_{s}}\,p_{\rm cm}(y-x)\bigg]\Big\}
 \label{amp-h-ll},
\end{eqnarray}
\begin{eqnarray}
{\cal A}_{h}^{LR}[C_{i}] &=& \frac{{\cal F}}{N_{C}} {\int}^{1}_{0} dx\,dy\,dz {\int}_{0}^{\infty} b db  \, b_{\phi}db_{\phi}\, b_{f} db_{f}\, {\alpha}_{s}(t_{h})\, H_{en}({\alpha}_{e},\, {\beta}_{h},\, b_{f},\, b_{\phi},\, b)\, E_{n}(t_{h})\,C_{i}(t_{h})\nonumber\\
&& \times \,2\,m_{B_{s}}^2\,
{\phi}_{-}\,\Big\{{\phi}_{B}^{a}(x)\,{\phi}_{\phi}^{v}(y)\,2\,p_{\rm cm}
\,\Big[x\,m_{B_{s}}-y\,E_{\phi}+(z-1)\,E_{w}\Big]\nonumber\\
&&  +\,{\phi}_{B}^{p}(x)\,r\,\bigg[{\phi}_{\phi}^{t}(y)\,m_{B_{s}}\,p_{\rm cm}(y-x)\nonumber\\
&& +\,{\phi}_{\phi}^{s}(y)\,
\big[E_{w}(x\,m_{B_{s}}-y\,E_{\phi})-y\,p_{\rm cm}^2+(z-1)\,w^2\big]\bigg]\Big\}
\label{amp-h-lr},
\end{eqnarray}
\begin{eqnarray}
{\cal A}_{h}^{SP}[C_{i}] &=& \frac{(-2){\cal F}}{N_{C}} {\int}^{1}_{0} dx\,dy\,dz {\int}_{0}^{\infty} b db  \, b_{\phi}db_{\phi}\, b_{f} db_{f}\, {\alpha}_{s}(t_{h})\, H_{en}({\alpha}_{e},\, {\beta}_{h},\, b_{f},\, b_{\phi},\, b)\, E_{n}(t_{h})\,C_{i}(t_{h})\nonumber\\
&& \times \,m_{B_{s}}\,w\,
\Big\{r\,m_{B_{s}}{\phi}_{B}^{a}(x)\,\bigg[({\phi}_{\phi}^{t}(y)\,{\phi}_{s}\,+\,{\phi}_{\phi}^{s}(y)\,{\phi}_{+})
\,p_{\rm cm}\,(1-y-z) \nonumber\\
&&+({\phi}_{\phi}^{t}(y)\,{\phi}_{+}\,+\,{\phi}_{\phi}^{s}(y)\,{\phi}_{s})
\,(y\,E_{\phi}-x\,m_{B_{s}}-z\,E_{w}+E_{w})\bigg]\nonumber\\
&& +\,{\phi}_{B}^{p}(x)\,{\phi}_{\phi}^{v}(y)\,\Big[{\phi}_{s}\,m_{B_{s}}\,p_{\rm cm}\,(x+z-1)\nonumber\\
&&+\,{\phi}_{+}\,[x\,m_{B_{s}}\,E_{\phi}-y\,m_{\phi}^2+(w^2
-m_{B_{s}}\,E_{w})(1-z)]\Big]\Big\}
\label{amp-h-sp},
\end{eqnarray}
\begin{eqnarray}
{\cal A}_{i}^{LL}[C_{i}] &=& {\cal F} {\int}^{1}_{0} dy\,dz {\int}_{0}^{\infty} b_{\phi}db_{\phi}\, b_{f} db_{f}\, {\alpha}_{s}(t_{i})\, H_{af}({\alpha}_{i},\, {\beta}_{i},\, b_{\phi},\, b_{f})\, E_{B}(t_{i})\,C_{i}(t_{i})\nonumber\\
&&\times\,2\,m_{B_{s}}^{2}\Big[ {\phi}_{-}\,{\phi}_{\phi}^{v}(y)\,m_{B_{s}}\,p_{\rm cm}\,(z\,r^2-r^2-z)\nonumber\\
&&-\,2\,r\,w\,{\phi}_{\phi}^{s}(y)\,[{\phi}_{s}\,(z\,E_{w}+E_{\phi})-\,{\phi}_{+}\,p_{\rm cm}\,(1-z)]
\Big]
\label{amp-i-ll},
\end{eqnarray}
\begin{equation}
{\cal A}_{i}^{LR} = {\cal A}_{i}^{LL}
\label{amp-i-lr},
\end{equation}
\begin{eqnarray}
{\cal A}_{i}^{SP}[C_{i}] &=& (-2){\cal F} {\int}^{1}_{0} dy\,dz {\int}_{0}^{\infty} b_{\phi}db_{\phi}\, b_{f} db_{f}\, H_{af}({\alpha}_{i},\, {\beta}_{i},\, b_{\phi},\, b_{f})\, E_{B}(t_{i})\,C_{i}(t_{i})\nonumber\\
&& \times\,2\,m_{B_{s}}\Big[w\,{\phi}_{\phi}^{v}(y)\,[{\phi}_{s}
z\,m_{B_{s}}\,p_{\rm cm}\,+
\,{\phi}_{+}\,(z\,w^2-m_{\phi}^2-z\,m_{B_{s}}\,E_{w})]\nonumber\\
&&
+\,{\phi}_{\phi}^{s}(y)\,{\phi}_{-}\,2\,r\,m_{B_{s}}\,(p_{\rm cm}^2+z\,w^2+E_{\phi}\,E_{w})
\Big]
\label{amp-i-sp},
\end{eqnarray}
\begin{eqnarray}
{\cal A}_{j}^{LL}[C_{i}] &=& {\cal F} {\int}^{1}_{0} dy\,dz {\int}_{0}^{\infty} b_{\phi}db_{\phi}\, b_{f} db_{f}\, {\alpha}_{s}(t_{j})\, H_{af}({\alpha}_{i},\, {\beta}_{j},\, b_{f},\, b_{\phi})\, E_{B}(t_{j})\,C_{i}(t_{j})\nonumber\\
&& \times\,2\,m_{B_{s}}\, \Big\{{\phi}_{-}\,{\phi}_{\phi}^{v}(y)\,p_{\rm cm}
\Big[(1-y)(m_{B_{s}}\,E_{\phi}+p_{\rm cm}^2+E_{\phi}\,E_{w})+w^2\Big] \nonumber\\
&&
-\,2\,r\,w\,m_{B_{s}}\,{\phi}_{s}\,\bigg[{\phi}_{\phi}^{t}(y)\,y\,p_{\rm cm}
-\,{\phi}_{\phi}^{s}(y)\,[E_{w}-E_{\phi}\,(y-1)]\bigg]\Big\}
\label{amp-j-ll},
\end{eqnarray}
\begin{equation}
{\cal A}_{j}^{LR}[C_{i}] = {\cal A}_{j}^{LL}[C_{i}]
\label{amp-j-lr},
\end{equation}
\begin{eqnarray}
{\cal A}_{j}^{SP}[C_{i}] &=& (-2){\cal F} {\int}^{1}_{0} dy\,dz {\int}_{0}^{\infty} b_{\phi}db_{\phi}\, b_{f} db_{f}\, {\alpha}_{s}(t_{j})\, H_{af}({\alpha}_{i},\, {\beta}_{j},\, b_{f},\, b_{\phi})\, E_{B}(t_{j})\,C_{i}(t_{j})\nonumber\\
&& \times\,2\,m_{B_{s}}^2\,\Big\{r\,{\phi}_{-}\,\bigg[{\phi}_{\phi}^{t}(y)
(y-1)\,m_{B_{s}}\,p_{\rm cm}
- {\phi}_{\phi}^{s}(y)\,[(y-1)\,p_{\rm cm}^2+(y-1)\,E_{\phi}\,E_{w}-w^2]
\nonumber\\
&&  +\,{\phi}_{s}\,{\phi}_{\phi}^{v}(y)2\,w\,p_{\rm cm}\Big\}
\label{amp-j-sp},
\end{eqnarray}
\begin{eqnarray}
{\cal A}_{k}^{LL}[C_{i}] &=& \frac{{\cal F}}{N_{C}} {\int}^{1}_{0} dx\,dy\,dz {\int}_{0}^{\infty} b db  \, b_{\phi}db_{\phi}\, b_{f} db_{f}\, {\alpha}_{s}(t_{k})\, H_{an}({\alpha}_{i},\, {\beta}_{k},\, b,\, b_{\phi},\, b_{f})\, E_{n}(t_{k})\,C_{i}(t_{k})\nonumber\\
&& \times \,2\,m_{B_{s}}^{2}
\Big\{{\phi}_{B}^{a}(x)\,\Big[{\phi}_{-}\,{\phi}_{\phi}^{v}(y)\,2\,E_{w}\,p_{\rm cm}\,
(x+z-1)\nonumber\\
&& +\,({\phi}_{s}\,{\phi}_{\phi}^{t}(y)\,+\,{\phi}_{+}\,{\phi}_{\phi}^{s}(y))\,r\,w\,p_{\rm cm}\,(y+z-1)\nonumber\\
&&
+\,({\phi}_{s}\,{\phi}_{\phi}^{s}(y)\,+\,{\phi}_{+}\,{\phi}_{\phi}^{t}(y))\,r\,w\,[(x-1)\,m_{B_{s}}-(y-1)\,E_{\phi}+z\,E_{w}]
\Big]\nonumber\\
&&+\,r_{b}\,m_{B_{s}}\,{\phi}_{B}^{p}(x)\,\Big[{\phi}_{-}\,{\phi}_{\phi}^{v}(y)\,p_{\rm cm}
+\,{\phi}_{s}\,{\phi}_{\phi}^{s}(y)\,2\,r\,w
\Big]\Big\}
\label{amp-k-ll},
\end{eqnarray}
\begin{eqnarray}
{\cal A}_{k}^{LR}[C_{i}] &=& \frac{{\cal F}}{N_{C}} {\int}^{1}_{0} dx\,dy\,dz {\int}_{0}^{\infty} b db  \, b_{\phi}db_{\phi}\, b_{f} db_{f}\, {\alpha}_{s}(t_{k})\, H_{an}({\alpha}_{i},\, {\beta}_{k},\, b,\, b_{\phi},\, b_{f})\, E_{n}(t_{k})\,C_{i}(t_{k})\nonumber\\
&& \times \,2\,m_{B_{s}}\,
\Big\{{\phi}_{B}^{a}(x)\,\Big[{\phi}_{-}\,{\phi}_{\phi}^{v}(y)\,2\,p_{\rm cm}\,
(x\,m_{B_{s}}\,E_{w}-y\,E_{w}\,E_{\phi}-y\,p_{\rm cm}^2-m_{\phi}^2+z\,w^2)\nonumber\\
&& +\,({\phi}_{s}\,{\phi}_{\phi}^{t}(y)\,+\,{\phi}_{+}\,{\phi}_{\phi}^{s}(y))
\,r\,w\,m_{B_{s}}\,p_{\rm cm}\,(1-y-z)\nonumber\\
&&
+\,({\phi}_{s}\,{\phi}_{\phi}^{s}(y)\,+\,{\phi}_{+}\,{\phi}_{\phi}^{t}(y))\,r\,w\,m_{B_{s}}
\,[(x-1)\,m_{B_{s}}-(y-1)\,E_{\phi}+z\,E_{w}]\nonumber\\
&&+\,r_{b}\,m_{B_{s}}^{2}
{\phi}_{B}^{p}(x)\,\Big[{\phi}_{-}\,{\phi}_{\phi}^{v}(y)\,p_{\rm cm}
+\,{\phi}_{s}\,{\phi}_{\phi}^{s}(y)\,2\,r\,w
\Big]\Big\}
\label{amp-k-lr},
\end{eqnarray}
\begin{eqnarray}
{\cal A}_{k}^{SP}[C_{i}] &=& \frac{(-2){\cal F}}{N_{C}} {\int}^{1}_{0} dx\,dy\,dz {\int}_{0}^{\infty} b db  \, b_{\phi}db_{\phi}\, b_{f} db_{f}\, {\alpha}_{s}(t_{k})\, H_{an}({\alpha}_{i},\, {\beta}_{k},\, b,\, b_{\phi},\, b_{f})\, E_{n}(t_{k})\,C_{i}(t_{k})\nonumber\\
&& \times \,m_{B_{s}}
\Big\{{\phi}_{B}^{a}(x)\,
\bigg[-r\,m_{B_{s}}^{2}\,{\phi}_{-}\,({\phi}_{\phi}^{t}(y)\,r_{b}\,p_{\rm cm}\,+\,{\phi}_{\phi}^{s}(y)\,E_{w})\nonumber\\
&&+\,r_{b}\,w\,m_{B_{s}}\,{\phi}_{\phi}^{v}(y)\,(
{\phi}_{s}\,p_{\rm cm}+\,{\phi}_{+}\,E_{\phi})\big]\nonumber\\
&& +\,
{\phi}_{B}^{p}(x)\bigg[r\,m_{B_{s}}\,{\phi}_{-}\,\big[
{\phi}_{\phi}^{t}(y)\,m_{B_{s}}\,p_{\rm cm}\,(x-y)\nonumber\\
&&-\,{\phi}_{\phi}^{s}(y)\,\big((1-x)\,m_{B_{s}}\,E_{w}-(1-y)\,E_{\phi}\,E_{w}-(1-y)\,p_{\rm cm}^2-z\,w^2\big)\big]
\nonumber\\
&&-\,w\,{\phi}_{\phi}^{v}(y)\big[
{\phi}_{s}\,m_{B_{s}}\,p_{\rm cm}\,(x+z-1)\,\nonumber\\
&&-\,
{\phi}_{+}\,\big((y-1)\,m_{\phi}^2
+(1-x)\,m_{B_{s}}\,E_{\phi}+z\,(w^2-m_{B_{s}}\,E_{w})\big)\big]\bigg]\Big\}
\label{amp-k-sp},
\end{eqnarray}
\begin{eqnarray}
{\cal A}_{l}^{LL}[C_{i}] &=& \frac{{\cal F}}{N_{C}} {\int}^{1}_{0} dx\,dy\,dz {\int}_{0}^{\infty} b db  \, b_{\phi}db_{\phi}\, b_{f} db_{f}\, {\alpha}_{s}(t_{l})\, H_{an}({\alpha}_{i},\, {\beta}_{l},\, b,\, b_{\phi},\, b_{f})\, E_{n}(t_{l})\,C_{i}(t_{l})\nonumber\\
&& \times \,2\,m_{B_{s}}
{\phi}_{B}^{a}(x)\,\Big\{{\phi}_{-}\,{\phi}_{\phi}^{v}(y)\,2\,p_{\rm cm}\,
\Big[x\,m_{B_{s}}\,E_{w}-z\,w^2+(y-1)(m_{B_{s}}\,E_{\phi}-m_{\phi}^{2})\Big]\nonumber\\
&& +\,r\,w\,m_{B_{s}}\,({\phi}_{s}\,{\phi}_{\phi}^{t}(y)\,+\,{\phi}_{+}\,{\phi}_{\phi}^{s}(y))p_{\rm cm}\,(y+z-1)\nonumber\\
&&
+\,r\,w\,m_{B_{s}}\,({\phi}_{s}\,{\phi}_{\phi}^{s}(y)+\,{\phi}_{+}\,{\phi}_{\phi}^{t}(y))
\,[x\,m_{B_{s}}+(y-1)\,E_{\phi}-z\,E_{w}]
\Big\}
\label{amp-l-ll},
\end{eqnarray}
\begin{eqnarray}
{\cal A}_{l}^{LR}[C_{i}] &=& \frac{{\cal F}}{N_{C}} {\int}^{1}_{0} dx\,dy\,dz {\int}_{0}^{\infty} b db  \, b_{\phi}db_{\phi}\, b_{f} db_{f}\, {\alpha}_{s}(t_{l})\, H_{an}({\alpha}_{i},\, {\beta}_{l},\, b,\, b_{\phi},\, b_{f})\, E_{n}(t_{l})\,C_{i}(t_{l})\nonumber\\
&& \times \,2\,m_{B_{s}}^2
{\phi}_{B}^{a}(x)\,\Big\{{\phi}_{-}\,{\phi}_{\phi}^{v}(y)\,2\,E_{w}\,p_{\rm cm}\,
(x-z)\nonumber\\
&& +\,r\,w\,p_{\rm cm}\,({\phi}_{s}\,{\phi}_{\phi}^{t}(y)\,+\,{\phi}_{+}\,{\phi}_{\phi}^{s}(y))\,(1-y-z)\nonumber\\
&&
+\,r\,w\,({\phi}_{s}\,{\phi}_{\phi}^{s}(y)\,+\,{\phi}_{+}\,{\phi}_{\phi}^{t}(y))\,[x\,m_{B_{s}}+(y-1)\,E_{\phi}-z\,E_{w}]\Big\} \label{amp-l-lr},
\end{eqnarray}
\begin{eqnarray}
{\cal A}_{l}^{SP}[C_{i}] &=& \frac{(-2){\cal F}}{N_{C}} {\int}^{1}_{0} dx\,dy\,dz {\int}_{0}^{\infty} b db  \, b_{\phi}db_{\phi}\, b_{f} db_{f}\, {\alpha}_{s}(t_{l})\, H_{an}({\alpha}_{i},\, {\beta}_{l},\, b,\, b_{\phi},\, b_{f})\, E_{n}(t_{l})\,C_{i}(t_{l})\nonumber\\
&& \times \,
{\phi}_{B}^{p}(x)\Big\{r\,m_{B_{s}}^2
{\phi}_{-}\,\bigg[{\phi}_{\phi}^{t}(y)\,m_{B_{s}}\,p_{\rm cm}\,(y+x-1)\nonumber\\
&&-\,{\phi}_{\phi}^{s}(y)\,[(1-y)\,E_{\phi}\,E_{w}-x\,m_{B_{s}}\,E_{w}-(y-1)\,p_{\rm cm}^2+z\,w^2]\bigg]
\nonumber\\
&&-\,w\,m_{B_{s}}\,{\phi}_{\phi}^{v}(y)\,\bigg[
{\phi}_{s}\,m_{B_{s}}\,p_{\rm cm}\,(x-z)\nonumber\\
&&+\,{\phi}_{+}\big[(y-1)\,m_{\phi}^2
+x\,m_{B_{s}}\,E_{\phi}+z\,(w^2-E_{w}\,m_{B_{s}})\Big]\bigg]\Big\}
\label{amp-l-lr},
\end{eqnarray}
\begin{eqnarray}
{\cal A}_{m}^{LL}[C_{i}] &=& -2\,{\cal F} dy\,dz {\int}_{0}^{\infty} b_{\phi}db_{\phi}\, b_{f} db_{f}\, {\alpha}_{s}(t_{m})\, H_{af}({\alpha}_{m},\, {\beta}_{m},\, b_{f},\, b_{\phi})\, E_{B}(t_{m})\,C_{i}(t_{m})\nonumber\\
&& \times\,
\Big\{m_{B_{s}}\,p_{\rm cm}\,{\phi}_{-}\,{\phi}_{\phi}^{v}(y)\Big[
y\,(E_{\phi}\,m_{B_{s}}+p_{\rm cm}^2+E_{\phi}\,E_{w})+w^2\Big]\nonumber\\
&& +2\,r\,w\,m_{B_{s}}^2\,{\phi}_{s}\,\Big[{\phi}_{\phi}^{t}(y)\,\,p_{\rm cm}\,(1-y)\,+\,
{\phi}_{\phi}^{s}(y)\,(y\,E_{\phi}+E_{w})\Big]
\Big\}
\label{amp-m-ll},
\end{eqnarray}
\begin{equation}
{\cal A}_{m}^{LR}[C_{i}] = {\cal A}_{m}^{LL}[C_{i}]
\label{amp-m-lr},
\end{equation}
\begin{eqnarray}
{\cal A}_{m}^{SP}[C_{i}] &=& 4{\cal F} {\int}^{1}_{0} dy\,dz {\int}_{0}^{\infty} b_{\phi}db_{\phi}\, b_{f} db_{f}\, {\alpha}_{s}(t_{m})\, H_{af}({\alpha}_{m},\, {\beta}_{m},\, b_{f},\, b_{\phi})\, E_{B}(t_{m})\,C_{i}(t_{m})\nonumber\\
&&\times\,m_{B_{s}}^{2} \Big\{
r\,{\phi}_{-}\,\bigg[{\phi}_{\phi}^{t}(y)\,y\,m_{B_{s}}\,p_{\rm cm}\,+\,{\phi}_{\phi}^{s}(y)\,
(y\,p_{\rm cm}^2+y\,E_{\phi}\,E_{w}+w^2)\bigg]
 \nonumber \\
&&
+\,{\phi}_{s}\,{\phi}_{\phi}^{v}(y)\,2\,w\,p_{\rm cm}
\Big\}
\label{amp-m-sp},
\end{eqnarray}
\begin{eqnarray}
{\cal A}_{n}^{LL}[C_{i}] &=& {\cal F} {\int}^{1}_{0} dy\,dz {\int}_{0}^{\infty} b_{\phi}db_{\phi}\, b_{f} db_{f}\, {\alpha}_{s}(t_{n})\, H_{af}({\alpha}_{m},\, {\beta}_{n},\, b_{f},\, b_{\phi})\, E_{B}(t_{n})\,C_{i}(t_{n})\nonumber\\
&& \times\,2\,m_{B_{s}}^{2}
\Big\{{\phi}_{-}\,{\phi}_{\phi}^{v}(y)\,m_{B_{s}}\,p_{\rm cm}(z\,r^2-z+1)\nonumber\\
&& +\,2\,r\,w\,{\phi}_{\phi}^{s}(y)\,\bigg[{\phi}_{s}\,[E_{\phi}-(z-1)\,E_{w}] +\,{\phi}_{+}\,z\,p_{\rm cm}\bigg]
\Big\}
\label{amp-n-ll},
\end{eqnarray}
\begin{equation}
{\cal A}_{n}^{LR}[C_{i}] = {\cal A}_{n}^{LL}[C_{i}]
\label{amp-n-lr},
\end{equation}
\begin{eqnarray}
{\cal A}_{n}^{SP}[C_{i}] &=& (-2){\cal F} {\int}^{1}_{0}dy\,dz {\int}_{0}^{\infty} b_{\phi}db_{\phi}\, b_{f} db_{f}\, {\alpha}_{s}(t_{n})\, H_{af}({\alpha}_{m},\, {\beta}_{n},\, b_{f},\, b_{\phi})\, E_{B}(t_{n})\,C_{i}(t_{n})\nonumber\\
&&\times\,2\,m_{B_{s}} \Big\{{\phi}_{-}\,{\phi}_{\phi}^{s}(y)\,2\,r\,m_{B_{s}}\,[(z-1)\,w^2-p_{\rm cm}^2-E_{\phi}\,E_{w}]\nonumber\\
&&
-\,w\,{\phi}_{\phi}^{v}(y)\,\bigg[{\phi}_{s}\,(1-z)\,m_{B_{s}}\,p_{\rm cm}\,-\,{\phi}_{+}\,[(1-z)(w^2-E_{w}\,m_{B_{s}})-m_{\phi}^2
]\bigg]
\Big\}
\label{amp-n-sp},
\end{eqnarray}
\begin{eqnarray}
{\cal A}_{o}^{LL}[C_{i}] &=& \frac{{\cal F}}{N_{C}} {\int}^{1}_{0} dx\,dy\,dz {\int}_{0}^{\infty} b db  \, b_{\phi}db_{\phi}\, b_{f} db_{f}\, {\alpha}_{s}(t_{o})\, H_{an}({\alpha}_{m},\, {\beta}_{o},\, b,\, b_{\phi},\, b_{f})\, E_{n}(t_{o})\,C_{i}(t_{o})\nonumber\\
&& \times \,2\,m_{B_{s}}
\Big\{{\phi}_{B}^{a}(x)\,\Big[{\phi}_{-}\,{\phi}_{\phi}^{v}(y)\,2\,p_{\rm cm}\,
\big[E_{w}\big((x-1)m_{B_{s}}+y\,E_{\phi}\big)+y\,p_{\rm cm}^2-(z-1)\,w^2\big]\nonumber\\
&&+\,r\,w\,m_{B_{s}}\,p_{\rm cm}\,({\phi}_{s}\,{\phi}_{\phi}^{t}(y)\,+\,{\phi}_{+}\,{\phi}_{\phi}^{s}(y))\,\,(1-y-z)\nonumber\\
&&
+\,r\,w\,m_{B_{s}}\,({\phi}_{s}\,{\phi}_{\phi}^{s}(y)\,+\,{\phi}_{+}\,{\phi}_{\phi}^{t}(y))\,[(x-1)\,m_{B_{s}}+y\,E_{\phi}-(z-1)\,E_{w}]
\nonumber\\
&&+\,r_{b}\,m_{B_{s}}^2\,{\phi}_{B}^{p}(x)\,\Big[{\phi}_{-}\,{\phi}_{\phi}^{v}(y)\,p_{\rm cm}\,+\,{\phi}_{s}\,{\phi}_{\phi}^{s}(y)\,2\,r\,w
\Big]\Big\}
\label{amp-o-ll},
\end{eqnarray}
\begin{eqnarray}
{\cal A}_{o}^{LR}[C_{i}] &=& \frac{{\cal F}}{N_{C}} {\int}^{1}_{0} dx\,dy\,dz {\int}_{0}^{\infty} b db  \, b_{\phi}db_{\phi}\, b_{f} db_{f}\, {\alpha}_{s}(t_{o})\, H_{an}({\alpha}_{m},\, {\beta}_{o},\, b,\, b_{\phi},\, b_{f})\, E_{n}(t_{o})\,C_{i}(t_{o})\nonumber\\
&& \times \,2\,m_{B_{s}}^2
\Big\{{\phi}_{B}^{a}(x)\,\Big[{\phi}_{-}\,{\phi}_{\phi}^{v}(y)\,2\,E_{w}\,p_{\rm cm}\,
(x-z)\nonumber\\
&& +\,r\,w\,p_{\rm cm}\,({\phi}_{s}\,{\phi}_{\phi}^{t}(y)\,+\,{\phi}_{+}\,{\phi}_{\phi}^{s}(y))\,(y+z-1)\nonumber\\
&&
+\,r\,w\,({\phi}_{s}\,{\phi}_{\phi}^{s}(y)\,+\,{\phi}_{+}\,{\phi}_{\phi}^{t}(y))\,[(x-1)\,m_{B_{s}}+y\,E_{\phi}-(z-1)\,E_{w}]
\Big]\nonumber\\
&&+\,r_{b}\,m_{B_{s}}\,{\phi}_{B}^{p}(x)\,\Big[{\phi}_{-}\,{\phi}_{\phi}^{v}(y)\,p_{\rm cm}\,+\,{\phi}_{s}\,{\phi}_{\phi}^{s}(y)\,2\,r\,w
\Big]\Big\}
\label{amp-o-lr},
\end{eqnarray}
\begin{eqnarray}
{\cal A}_{o}^{SP}[C_{i}] &=& \frac{(-2){\cal F}}{N_{C}} {\int}^{1}_{0} dx\,dy\,dz {\int}_{0}^{\infty} b db  \, b_{\phi}db_{\phi}\, b_{f} db_{f}\, {\alpha}_{s}(t_{o})\, H_{an}({\alpha}_{m},\, {\beta}_{o},\, b,\, b_{\phi},\, b_{f})\, E_{n}(t_{o})\,C_{i}(t_{o})\nonumber\\
&& \times \,m_{B_{s}}\,
\Big\{{\phi}_{B}^{a}(x)\,
\bigg[r\,r_{b}\,m_{B_{s}}^{2}\,{\phi}_{-}\,({\phi}_{\phi}^{t}(y)\,p_{\rm cm}\,-\,{\phi}_{\phi}^{s}(y)\,E_{w})
\nonumber\\
&&+\,r_{b}\,w\,m_{B_{s}}\,{\phi}_{\phi}^{v}(y)\,({\phi}_{s}\,p_{\rm cm}\,-\,
{\phi}_{+}\,E_{\phi})\bigg]\nonumber\\
&& +\,
{\phi}_{B}^{p}(x)\Big[r\,m_{B_{s}}\,{\phi}_{-}\,[\,{\phi}_{\phi}^{t}(y)\,m_{B_{s}}\,p_{\rm cm}\,(1-y-x)\nonumber\\
&&\,-\,
{\phi}_{\phi}^{s}(y)\,
\big((z-1)\,w^2-y\,p_{\rm cm}^2+(1-x)\,m_{B_{s}}\,E_{w}-y\,E_{\phi}\,E_{w}\big)]
\nonumber\\
&&-\,w\,
{\phi}_{\phi}^{v}(y)\,[{\phi}_{s}\,m_{B_{s}}\,p_{\rm cm}\,(x-z)\nonumber\\
&&+\,
{\phi}_{+}\,\big((1-y)\,m_{\phi}^2
-x\,m_{B_{s}}\,E_{\phi}+z\,(m_{B_{s}}\,E_{w}-w^2)\big)]\Big]\Big\}
\label{amp-o-sp},
\end{eqnarray}
\begin{eqnarray}
{\cal A}_{p}^{LL}[C_{i}] &=& \frac{{\cal F}}{N_{C}} {\int}^{1}_{0} dx\,dy\,dz {\int}_{0}^{\infty} b db  \, b_{\phi}db_{\phi}\, b_{f} db_{f}\, {\alpha}_{s}(t_{p})\, H_{an}({\alpha}_{m},\, {\beta}_{p},\, b,\, b_{\phi},\, b_{f})\, E_{n}(t_{p})\,C_{i}(t_{p})\nonumber\\
&& \times \,2\,m_{B_{s}}^2\,
{\phi}_{B}^{a}(x)\,\Big\{2\,E_{w}\,p_{\rm cm}\,{\phi}_{-}\,{\phi}_{\phi}^{v}(y)\,
(x+z-1)\nonumber\\
&&+\,r\,w\,p_{\rm cm}\,({\phi}_{s}\,{\phi}_{\phi}^{t}(y)\,+\,{\phi}_{+}\,{\phi}_{\phi}^{s}(y))\,(1-y-z)\nonumber\\
&&+\,r\,w\,({\phi}_{s}\,{\phi}_{\phi}^{s}(y)\,+\,{\phi}_{+}\,{\phi}_{\phi}^{t}(y))
\,[x\,m_{B_{s}}-y\,E_{\phi}+(z-1)\,E_{w}]
\Big\}
\label{amp-p-ll},
\end{eqnarray}
\begin{eqnarray}
{\cal A}_{p}^{LR}[C_{i}] &=& \frac{{\cal F}}{N_{C}} {\int}^{1}_{0} dx\,dy\,dz {\int}_{0}^{\infty} b db  \, b_{\phi}db_{\phi}\, b_{f} db_{f}\, {\alpha}_{s}(t_{p})\, H_{an}({\alpha}_{m},\, {\beta}_{p},\, b,\, b_{\phi},\, b_{f})\, E_{n}(t_{p})\,C_{i}(t_{p})\nonumber\\
&& \times \,2\,m_{B_{s}}\,
{\phi}_{B}^{a}(x)\,\Big\{2\,p_{\rm cm}\,{\phi}_{-}\,{\phi}_{\phi}^{v}(y)\,
\big[E_{w}(x\,m_{B_{s}}-E_{\phi}\,y)-y\,p_{\rm cm}^2+(z-1)w^2\big]\nonumber\\
&& +\,r\,w\,p_{\rm cm}\,m_{B_{s}}\,({\phi}_{s}\,{\phi}_{\phi}^{t}(y)\,+\,{\phi}_{+}\,{\phi}_{\phi}^{s}(y))\,(y+z-1)\nonumber\\
&&+\,r\,w\,m_{B_{s}}\,({\phi}_{s}\,{\phi}_{\phi}^{s}(y)\,+\,{\phi}_{+}\,{\phi}_{\phi}^{t}(y))
\,[x\,m_{B_{s}}-y\,E_{\phi}+(z-1)\,E_{w}]\Big\}
\label{amp-p-lr},
\end{eqnarray}
\begin{eqnarray}
{\cal A}_{p}^{SP}[C_{i}] &=& \frac{(-2){\cal F}}{N_{C}} {\int}^{1}_{0} dx\,dy\,dz {\int}_{0}^{\infty} b db  \, b_{\phi}db_{\phi}\, b_{f} db_{f}\, {\alpha}_{s}(t_{p})\, H_{an}({\alpha}_{m},\, {\beta}_{p},\, b,\, b_{\phi},\, b_{f})\, E_{n}(t_{p})\,C_{i}(t_{p})\nonumber\\
&& \times \,m_{B_{s}}\,
{\phi}_{B}^{p}(x)\Big\{r\,m_{B_{s}}\,
{\phi}_{-}\,\bigg[m_{B_{s}}\,p_{\rm cm}\,{\phi}_{\phi}^{t}(y)\,(y-x)\nonumber\\
&&-\,{\phi}_{\phi}^{s}(y)\,
[y\,E_{\phi}\,E_{w}-x\,m_{B_{s}}\,E_{w}+y\,p_{\rm cm}^2-(z-1)\,w^2]\bigg]
\nonumber\\
&&-\,w\,
{\phi}_{\phi}^{v}(y)\,\bigg[m_{B_{s}}\,p_{\rm cm}\,{\phi}_{s}\,(x+z-1)\nonumber\\
&&+\,
{\phi}_{+}\,\big[y\,m_{\phi}^2
-x\,m_{B_{s}}\,E_{\phi}+(1-z)(m_{B_{s}}\,E_{w}-w^2)\big]\bigg]\Big\}
\label{amp-p-sp},
\end{eqnarray}
  where the color number $N_{c}$ $=$ $3$ and the color factor $C_{F}$ $=$ $4/3$.
  The superscript ${\rho}$ of the amplitude building block ${\cal A}_{\sigma}^{\rho}$
  refers to the three possible Dirac structures ${\Gamma}_{1}{\otimes}{\Gamma}_{2}$
  of the operators $(\bar{q}_{1}q_{2})_{{\Gamma}_{1}}(\bar{q}_{3}q_{4})_{{\Gamma}_{2}}$,
  namely ${\rho}$ $=$ $LL$ for $(V-A){\otimes}(V-A)$, ${\rho}$ $=$ $LR$
  for $(V-A){\otimes}(V+A)$ and ${\rho}$ $=$ $SP$ for $-2(S-P){\otimes}(S+P)$.
  The subscript ${\sigma}$ of ${\cal A}_{\sigma}^{\rho}$ (${\sigma}$ $=$ $a$
  $b$, ${\cdots}$, $p$) corresponds to the sub-diagram indices of Fig.\ref{fey}.
  The variables of $b$, $b_{\phi}$, $b_{f}$ are the conjugate variables of
  the transverse momentum $p_{T}$, $k_{T}$, $l_{T}$, respectively.

 The function $H_{i}$ and Sudakov factor $E_{i}$ are defined as
  \begin{eqnarray}
  H_{ef}({\alpha},{\beta},b_{i},b_{j}) &=&
     K_{0}(b_{i}\sqrt{-{\alpha}}) \Big\{
    {\theta}(b_{i}-b_{j})\,K_{0}(b_{i}\sqrt{-{\beta}})\,I_{0}(b_{j}\sqrt{-{\beta}})
          + (b_{i}{\leftrightarrow}b_{j}) \Big\}
  \label{hef}, \\
  H_{en}({\alpha},{\beta},b_{i},b_{j},b_{k}) &=&
     \Big\{ {\theta}(-{\beta})\,K_{0}(b_{i}\sqrt{-{\beta}})
          +  \frac{\pi}{2}{\theta}({\beta}) \Big[
          i\,J_{0}(b_{i}\sqrt{\beta})-Y_{0}(b_{i}\sqrt{\beta}) \Big] \Big\}
  \nonumber \\ &{\times}&
     \Big\{ {\theta}(b_{i}-b_{j})\,K_{0}(b_{i}\sqrt{-{\alpha}})\,I_{0}(b_{j}\sqrt{-{\alpha}})
          + (b_{i}{\leftrightarrow}b_{j}) \Big\} {\delta}(b_{j}-b_{k})
  \label{hen}, \\
  H_{af}({\alpha},{\beta},b_{i},b_{j}) &=&
     \Big\{ {\theta}(b_{i}-b_{j}) \Big[ i\,J_{0}(b_{i}\sqrt{\beta})
       - Y_{0}(b_{i}\sqrt{\beta}) \Big] J_{0}(b_{j}\sqrt{\beta})
          + (b_{i}{\leftrightarrow}b_{j}) \Big\}
  \nonumber \\ &{\times}&
     \frac{{\pi}^{2}}{4} \Big\{ i\,J_{0}(b_{i}\sqrt{\alpha})
                                -  Y_{0}(b_{i}\sqrt{\alpha}) \Big\}
  \label{haf}, \\
  H_{an}({\alpha},{\beta},b_{i},b_{j},b_{k}) &=&
      \frac{\pi}{2} \Big\{ {\theta}(-{\beta})K_{0}(b_{i}\sqrt{-{\beta}})
     +\frac{\pi}{2} {\theta}({\beta}) \Big[ i\,J_{0}(b_{i}\sqrt{\beta})
     -Y_{0}(b_{i}\sqrt{\beta}) \Big] \Big\}
  \nonumber \\ & & \!\!\!\!\!\!\!\!\!\!\!\!\!\!\!\!
           \!\!\!\!\!\!\!\!\!\!\!\!\!\!\!\!\!\!\!\! {\times}\,
     \Big\{ {\theta}(b_{i}-b_{j}) \Big[ i\,J_{0}(b_{i}\sqrt{\alpha})
       - Y_{0}(b_{i}\sqrt{\alpha})\Big] J_{0}(b_{j}\sqrt{\alpha})
     + (b_{i}{\leftrightarrow}b_{j}) \Big\} {\delta}(b_{j}-b_{k})
  \label{han},
  \end{eqnarray}
  \begin{eqnarray}
  E_{\phi}(t) &=& {\exp} \Big\{-S_{B_{s}}(t)-S_{f_{0}}(t)\Big\}
  \label{e-phi}, \\
  E_{f}(t) &=& {\exp}\Big\{-S_{B_{s}}(t)-S_{\phi}(t)\Big\}
  \label{e-f}, \\
  E_{B}(t) &=& {\exp}\Big\{-S_{f_{0}}(t)-S_{\phi}(t)\Big\}
  \label{e-b}, \\
  E_{n}(t) &=& {\exp}\Big\{-S_{B_{s}}(t)-S_{f_{0}}(t)-S_{\phi}(t)\Big\}
  \label{e-n},
  \end{eqnarray}
  \begin{eqnarray}
  S_{B_{s}}(t) &=& s(x,b,p_{B_{s}}^{+})
      +2{\int}_{1/b}^{t}\,\frac{d{\mu}}{\mu}\,{\gamma}_{q}
  \label{s-b}, \\
  S_{\phi}(t) &=& s(y,b_{\phi},p_{\phi}^{+})
      +s(\bar{y},b_{\phi},p_{\phi}^{+})
      +2{\int}_{1/b_{\phi}}^{t}\,\frac{d{\mu}}{\mu}\,{\gamma}_{q}
  \label{s-phi}, \\
  S_{f_{0}}(t) &=& s(z,b_{f},q^{+})+s(\bar{z},b_{f},q^{+})
      +2{\int}_{1/b_{f}}^{t}\,\frac{d{\mu}}{\mu}\,{\gamma}_{q}
  \label{s-f},
  \end{eqnarray}
 where the subscripts $i$ $=$ $ef$, $en$, $af$, $an$ of the function $H_{i}$
 correspond to the factorizable emission topologies, the nonfactorizable e
 mission topologies, the factorizable annihilation topologies, and the
 nonfactorizable annihilation topologies, respectively.
 $I_{0}$, $J_{0}$, $K_{0}$ and $Y_{0}$ are the Bessel functions.
 The expression of $s(x,b,Q)$ can be found in of Ref.\cite{prd52.3958}.
 ${\gamma}_{q}$ $=$ $-{\alpha}_{s}/{\pi}$ is the quark anomalous dimension.

 The parameters of ${\alpha}_{i}$ and ${\beta}_{i}$ are the virtualities of
 gluons and quarks. The subscript $i$ of ${\alpha}_{i}$, ${\beta}_{i}$, $t_{i}$
 corresponds to the indices of Fig.\ref{fey}. The explicit definitions of the
 virtualities and typical scale $t_{i}$ are given as follows.
  \begin{eqnarray}
 {\alpha}_{a} &=& x^2\,m_{B_{s}}^2+z^2\,w^2-2\,x\,z\,m_{B_{s}}\,E_{w}
  \label{alpha-a}, \\
 {\alpha}_{e} &=& x^2\,m_{B_{s}}^2+y^2\,m_{\phi}^2-2\,x\,y\,m_{B_{s}}\,E_{\phi}
  \label{alpha-e}, \\
 {\alpha}_{i} &=& \bar{y}^2\,m_{\phi}^2+z^2\,w^2+\bar{y}\,z\,(m_{B_{s}}^2-m_{\phi}^2-w^2)
  \label{alpha-i}, \\
 {\alpha}_{m} &=& y^2\,m_{\phi}^2+\bar{z}^2\,w^2+ y\,\bar{z}(m_{B_{s}}^2-m_{\phi}^2-w^2)
  \label{alpha-m}, \\
 {\beta}_{a} &=& (1-r_{b}^2)\,m_{B_{s}}^2+z^2\,w^2-2\,z\,m_{B_{s}}\,E_{w}
  \label{beta-a}, \\
 {\beta}_{b} &=& w^2+x^2\,m_{B_{s}}^2-2\,x\,m_{B_{s}}\,E_{w}
  \label{beta-b}, \\
 {\beta}_{c}   &=& (x-y)\,(x-z)\,m_{B_{s}}^{2}
  \nonumber \\ &+& (y-z)\,(y-x)\,m_{\phi}^{2}
  \nonumber \\ &+& (z-x)\,(z-y)\,w^{2}
  \label{beta-c}, \\
 {\beta}_{d} &=& \left.{\beta}_{c}\right\vert_{y{\to}\bar{y}}
  \label{beta-d}, \\
 {\beta}_{e} &=& (1-r_{b}^2)\,m_{B_{s}}^2+y^2\,m_{\phi}^{2}-2\,y\,m_{B_{s}}\,E_{\phi}
  \label{beta-e}, \\
 {\beta}_{f} &=& m_{\phi}^2+x^2\,m_{B_{s}}^2-2\,x\,m_{B_{s}}\,E_{\phi}
  \label{beta-f}, \\
 {\beta}_{g} &=& {\beta}_{c}
  \label{beta-g}, \\
 {\beta}_{h} &=& \left.{\beta}_{g}\right\vert_{z{\to}\bar{z}}
  \label{beta-h}, \\
 {\beta}_{i} &=& m_{\phi}^{2}+z^2\,w^2+z\,(m_{B_{s}}^2-m_{\phi}^2-w^2)
  \label{beta-i}, \\
 {\beta}_{j} &=& \bar{y}^2\,m_{\phi}^2+w^2+\bar{y}\,(m_{B_{s}}^2-m_{\phi}^2-w^2)
  \label{beta-j}, \\
 {\beta}_{k} &=& \left.{\beta}_{c}\right\vert^{x{\to}\bar{x}}_{y{\to}\bar{y}}
              -m_{b}^2
  \label{beta-k}, \\
 {\beta}_{l} &=& \left.{\beta}_{c}\right\vert_{y{\to}\bar{y}}
  \label{beta-l}, \\
 {\beta}_{m} &=& w^2+y^2\,m_{\phi}^2+y\,(m_{B_{s}}^2-m_{\phi}^2-w^2)
  \label{beta-l}, \\
 {\beta}_{n} &=& \bar{z}^2\,w^2+m_{\phi}^2+\bar{z}\,(m_{B_{s}}^2-m_{\phi}^2-w^2)
  \label{beta-n}, \\
 {\beta}_{o} &=& \left.{\beta}_{c}\right\vert^{x{\to}\bar{x}}_{z{\to}\bar{z}}
              -m_{b}^2
  \label{beta-o}, \\
 {\beta}_{p} &=&  \left.{\beta}_{c}\right\vert_{z{\to}\bar{z}}
  \label{beta-p},
  \end{eqnarray}
  \begin{eqnarray}
  t_{a,b} &=& {\max}\{\sqrt{-{\alpha}_{a}},\sqrt{{\vert}{\beta}_{a,b}{\vert}},1/b,1/b_{f}\}
  \label{t-ab}, \\
  t_{c,d} &=& {\max}\{\sqrt{-{\alpha}_{a}},\sqrt{{\vert}{\beta}_{c,d}{\vert}},1/b,1/b_{\phi}\}
  \label{t-cd}, \\
  t_{e,f} &=& {\max}\{\sqrt{-{\alpha}_{e}},\sqrt{{\vert}{\beta}_{e,f}{\vert}},1/b,1/b_{\phi}\}
  \label{t-ef}, \\
  t_{g,h} &=& {\max}\{\sqrt{-{\alpha}_{e}},\sqrt{{\vert}{\beta}_{g,h}{\vert}},1/b,1/b_{f}\}
  \label{t-gh}, \\
  t_{i,j} &=& {\max}\{\sqrt{{\alpha}_{i}},\sqrt{{\vert}{\beta}_{i,j}{\vert}},1/b_{\phi},1/b_{f}\}
  \label{t-ij}, \\
  t_{k,l} &=& {\max}\{\sqrt{{\alpha}_{i}},\sqrt{{\vert}{\beta}_{k,l}{\vert}},1/b,1/b_{f}\}
  \label{t-kl}, \\
  t_{m,n} &=& {\max}\{\sqrt{{\alpha}_{m}},\sqrt{{\vert}{\beta}_{m,n}{\vert}},1/b_{\phi},1/b_{f}\}
  \label{t-mn}, \\
  t_{o,p} &=& {\max}\{\sqrt{{\alpha}_{m}},\sqrt{{\vert}{\beta}_{o,p}{\vert}},1/b,1/b_{f}\}
  \label{t-op}.
  \end{eqnarray}
  \end{appendix}

 
 \end{document}